\documentclass{pazh}
\usepackage{graphicx}
\usepackage{latexsym}

\newcommand{\km}{\,\mbox{km}\,\mbox{s}^{-1}}
\def\Ha{\hbox{H$_\alpha$\,}}
\def\Hb{\hbox{H$_\beta$\,}}

\def\farcs{\hbox{$.\!\!^{\prime\prime}$}}

\def\Rc{$\mbox{R}_c$\,}

\begin{document}

\title{Observational evidence for AGN fueling. I. The merging of NGC~6104 with a companion.}

\author{A. A. Smirnova, A. V. Moiseev, and V. L. Afanasiev}

\institute{Special Astrophysical Observatory, RAS, Nizhnii Arkhyz,
Karachaevo-Cherkesia, 357147 Russia }

\offprints{A.A.  Smirnova, \email{alexiya@sao.ru}}

\date{}

\titlerunning{Observational evidence for AGN fueling. I.  NGC 6104}

\authorrunning{Smirnova et al. }

\abstract{ We investigate in details the kinematics and morphology
of the Seyfert galaxy NGC~6104 in order to identify the mechanism
of gas transportation to the active galactic nucleus (AGN). Our
observational data were obtained at the 6-m Special Astrophysical
Observatory telescope with the MPFS integral-field spectrograph
and the SCORPIO universal device in three modes: direct imaging, a
scanning Fabry-Perot interferometer, and long-slit spectroscopy.
Images from the HST archive were invoked to study the structure of
the circumnuclear region. An analysis of deep images has revealed
that NGC~6104 is in the phase of active merging with a companion
galaxy. We have been able to study the detailed picture of ionized
gas motions up to galactocentric distances of 14 kpc and to
construct the stellar velocity field for the inner region. The
radial gas motions toward the AGN along the central bar play a
significant role at galactocentric distances of 1-5 kpc. In
addition, we have detected an outflow of ionized gas from the
nucleus that presumably resulted from the intrusion of a radio jet
into the ambient interstellar medium. Using diagnostic diagrams,
we estimate the contributions from the AGN and star formation to
the galactic gas ionization. We estimate the bar pattern speed by
the Tremaine-Weinberg method and show that the inner ring observed
in the galactic images has a resonant nature. Two possible ring
formation scenarios (before and during the interaction with a
companion) are discussed.}

\maketitle

\section{Introduction.}

The problem of galactic activity still remains to be far from
understood. Why are the nuclei active in some galaxies and
inactive in other ones? What is this, a short phase in the
lifetime of any galaxy or a fundamental difference of the nuclei?
There is no unequivocal answer to all these questions as yet. The
mechanism of gas transportation to the region controlled by the
gravity of the central object (a supermassive black hole, in the
common opinion) remains a main problem. More specifically, much of
the angular momentum must be drained from the gas that was
initially at galactocentric distances of several kpc to ensure its
motion toward the nucleus (see the reviews of Combes 2001; Wada
2004). Numerous attempts to associate the presence of an active
(Seyfert) galactic nucleus (AGN) with properties of the host
galaxy, such as the presence of a central bar or even a double bar
(Laine et al. 2002; Knapen 2005) or an inner minispiral (Martini
et al. 2003), have been made over the past decades.

Many authors have tried to find a correlation between the presence
of an AGN in a galaxy and its environment: the presence of
companions or traces of interaction (Dahari 1985; de Robertis et
al. 1998; Schmitt 2001; Knapen 2005). However, statistically
significant correlation of the properties of host galaxy  with
activity wasn't found in any of the listed papers. Kauffmann et
al. (2004) and Hall and Richards (2004) compiled various samples
of galaxies with active and normal (quiescent) nuclei from the
Sloan Digital Sky Survey and compared their morphological
properties. Again, no statistically significant differences
between galaxies with active and normal nuclei were found.

In a recent paper devoted to this problem (Martini 2004), it has
been suggested that the mechanism of excitation and maintenance of
activity could be complex and depend on several factors at once.
The paper by Garcia-Burillo et al. (2005), in which the authors
attempted to directly measure the molecular gas inflow rate into
an AGN, is of considerable interest in this respect. A detailed
analysis of the dynamics of inner regions in four Seyfert galaxies
leads to a paradoxical conclusion -- despite the presence of a
bar, the gravitational potential distribution in these galaxies
currently  prevents the gas transportation into the innermost
(less than 100-200 pc) region. Authors suggested several possible
solutions of this paradox: allowance for viscosity and tidal
friction that take away the angular momentum of gaseous clouds or
a periodic action of some additional factors help the gas reach
the nucleus (a mass redistribution in barred galaxies). In any
case, it is clear that we can understand why the "fuel" reaches
the region of gravitational influence of the central engine of
active galaxies only through a detailed analysis of the kinematics
and dynamics of inner (100-1000 pc) galactic regions.

With this paper, we begin a series of papers devoted to
observational evidences of gas inflow into the center, which must
ensure the (possibly periodic) existence of an AGN. Integral-field
spectroscopy allows us to trace the behavior of gas and stars both
in the immediate vicinity of an AGN and far from the center. In
addition, the method of ionization diagrams means to determination
the type of an ionization source in various galactic regions.

Here, using the galaxy NGC~6104 as an example, we wish to
trace how the interaction between galaxies affects the activity of
their nuclei. On the other hand, it is interesting to consider
the relationship between nuclear activity and peculiarities of
the motion of gas and stars in the galactic disk.

NGC~6104 (its basic parameters are given in Table 1) is interesting
in that no candidates for the role of a perturbing companion have
ever been found (Rafanelli et al. 1995), although some of the
authors pointed out its peculiar morphology (Jimenez-Benito et al.
2000; Malkan et al. 1998; Tomita et al. 1999). Only our deep
images have revealed outer asymmetric structures of low surface
brightness that can be unambiguously interpreted as a result of a
recent interaction with a companion that was most likely followed
by its disruption. Therefore, it is interesting to know whether
the radial gas flows toward the galactic center that we detected
are related to these signs of interaction, or not.

\begin{table}
\centering \caption{Main characteristics of NGC~6104} \label{t1}
\begin{tabular}{@{}llll@{}}
\hline
Type of activity (from NED)             &    Sy1.5        \\
Morphological type (from NED)           &   S(R)Pec       \\
Morphological type (from LEDA)          &     SBab         \\
Systemic velocity  (from NED)           &    8429  km/s   \\
Absolute magnitude$^1$                            &   -21.14         \\
Adopted distance$^2$                              &   113  Mpc        \\
Spatial scale                                     &   0.55  kpc$/''$    \\
\hline
\multicolumn{3}{l}{$^1$  magnitude in the B-band was taken from}\\
\multicolumn{3}{l}{~~~Ho \& Peng (2001) and given at distance 113 Mpc}\\
\multicolumn{3}{l}{$^2$  according to Mouri et al. (2002)}\\
\end{tabular}
\end{table}

\begin{table*}

 \caption{Log of observations NGC~6104 and its
environment} \label{t2}
\begin{tabular}{@{}llllll@{}}
\hline
    Date     &   Instrument            & Exposition & spectral   & spectral   & seeing   \\
             &                         &  (sec)     &  range     & resolution &          \\
\hline
2002 May  15 &   MPFS                  &   3 x 900  &    4600 - 7000\AA     &      8\AA       &  $1.5''$   \\
2002 Aug. 13 &   MPFS                  &   2 x 1200 &    4600 - 7000\AA     &      8\AA       &  $1.5''$   \\
\hline
2003 Dec. 24 &  SCORPIO (FPI)          &  32 x  90  &    $H_{\alpha}$       &      2.5\AA     &  $1.6''$   \\
2004 Jan. 31 &  SCORPIO                &   3 x 300  &         B             &                 &  $1.6''$   \\
             & (direct images)         &   3 x 180  &         V             &                 &  $1.6''$   \\
             &                         &   8 x 120  &         R             &                 &  $1.5''$   \\
2004 June 21 &  SCORPIO                &   4 x 300  &    4000 - 9000\AA     &     10\AA       &  $1.1''$   \\
             &          (long slit)    &            &                       &                 &            \\
2005 May 17  &  SCORPIO                &   2 x 900  &    4020 - 5600\AA     &      5\AA       &  $1.6''$   \\
             &          (long slit)    &            &                       &                 &            \\
 \hline
\end{tabular}
 \end{table*}

\section{Observations and data reduction.}

All of the observational data were obtained at the prime focus of
the 6-m Special Astrophysical Observatory (SAO) telescope. The
central region of the galaxy was observed with the multipupil
fiber spectrograph (MPFS). In order to obtain information about
the outer parts of the galaxy and its surroundings, SCORPIO focal
reducer was used in the following observational  modes: a scanning
Fabry-Perot interferometer (FPI), long-slit spectroscopy, and
direct broad-band imaging.  The CCD detectors were TK1024
($1024\times 1024$) and EEV42-40 ($2048\times2048$ pixels) in 2002
and 2003-2005, respectively. A log of observations is given in
Table 2.

\subsection{MPFS}

 MPFS (Afanasiev et al. 2001) can simultaneously take spectra from
240 spatial elements arranged in the form of a $16\times15$ lens
array with a scale of $1''$/element. The spectral range  included
both numerous emission lines of ionized gas ([OI]$\lambda6300$,
[OIII]$\lambda4959+5007$, [NII]$\lambda6548+6583$,
[SII]$\lambda6717+6731$, the Balmer \Ha and \Hb lines) and
absorption features characteristic of the galactic old stellar
population. We used a software package running in the IDL
environment to reduce the observational data. The MPFS data
reduction steps were briefly described by Moiseev et al. (2004).
The data reduction result is a data cube in which a spectrum
corresponds to each image pixel. NGC~6104 was observed twice in
the same spectral range. After the primary data reduction, the two
cubes were combined. The resulting field of view of the cube was
$22\times20''$ , while the spatial resolution after the smoothing
procedures corresponded to a seeing of $\sim2''$. The spectra of
spectrophotometric standard stars were used in order to transform
the galaxy's spectra to energy units.

The stellar velocity fields were constructed using a
cross-correlation technique modified to work with MPFS data
(Moiseev 2001). Because of the relatively low spectral resolution,
we could not study the distribution of the stellar velocity
dispersion within the field of view and can only conclude that it
does not exceed $200 \km$.

 By fitting the profiles of the
main emission lines with Gaussians, we constructed their surface
brightness distributions and radial velocity fields. The maps for
the most intense lines are shown in Figs. 1 and 2. The influence
of stellar absorption features,  primarily in the spectral region
with  hydrogen lines, constitutes a main problem in analyzing the
spectral lines of ionized gas. We used the following simple
technique that allows the contribution from the stellar continuum
to be taken into account, at least to a first approximation,
without using complex and not always unambiguous methods like
synthetic spectra. The absorption features under the \Ha and \Hb
lines were fitted with Gaussians, the positions of whose centers
were determined from the stellar velocity field (strictly
speaking, from its fit by the model of circular rotation), FWHM
was fixed (because of the low spectral resolution, we may ignore
the stellar velocity dispersion variations over the field), and
the amplitude was proportional to the continuum level\footnote{To
take into account the contribution from the AGN-related
non-thermal component, we subtracted the image of a point source
from the continuum maps in the innermost regions. The  resulting
maps corresponding to the stellar component alone were used for
calculation of the equivalent widths.} (i.e., corresponded to a
constant equivalent width). The free Gaussian parameters (the
absorption line FWHM and equivalent widths) were chosen by fitting
the spectra at the edge of the field of view, where the
contribution from the emission lines decreases and the absorption
features under \Ha and \Hb can be clearly detected.

\subsection{SCORPIO focal reducer.}

SCORPIO universal instrument (Afanasiev and Moiseev 2005) allows
various types of spectroscopic and photometric observations to be
performed with $\sim6'$ field of view. Below, we describe in
details each of the modes used.

\subsubsection{Fabry-Perot interferometer.}

The observations with the scanning FPI were performed in the \Ha
emission line; the spectral interval between the neighboring
interference orders was $\sim 28$ \AA. The required spectral range
was separated out by a narrow-band filter with FWHM of $\sim15$
\AA. During the observations, we sequentially took 32
interferograms of the object by changing the FPI plate gap. The
data reduction technique was described by Moiseev (2002a). The
data reduction result is a data cube in which a 32-channel
spectrum with a sampling step of 0.9\AA ~corresponds to each pixel
with the scale 1pix=0\farcs71. The radial velocity field and the
monochromatic \Ha image were constructed by fitting the emission
line profile with a Gaussian. Final spatial resolution after the
smoothing procedures corresponds to a seeing of 2\farcs1. The \Ha
flux map was calibrated in absolute energy units ($erg\,s^{-1}\,
cm^{-2}\,acrsec^{-2}$) by comparison with a similar MPFS map for the galactic
center.

\subsubsection{Direct imaging.}

We obtained images of NGC~6104 in the Johnson-Causins B, V, \Rc
bands with a scale of 0\farcs35/pixel. The data reduction was
performed in IDL and included a number of standard procedures:
bias subtraction, flat fielding, and cosmic ray particle hits
removal. The magnitude calibration was performed using secondary
standards. For this purpose, we obtained short-exposure images of
both the galaxy itself and fields with photometric standards using
the Zeiss-1000 SAO telescope. Having calibrated these images using
a standard technique, we were able to perform the photometric
calibration of our deep images. The achieved surface brightness
levels (for a signal-to-noise ratio of 3) were 26.5, 25.4, and
25.5$^m/\Box''$ for the B, V, and \Rc ~images, respectively.
\begin{figure}
\hspace{0cm}(a) \\
\centerline{\includegraphics[width=8.5cm]{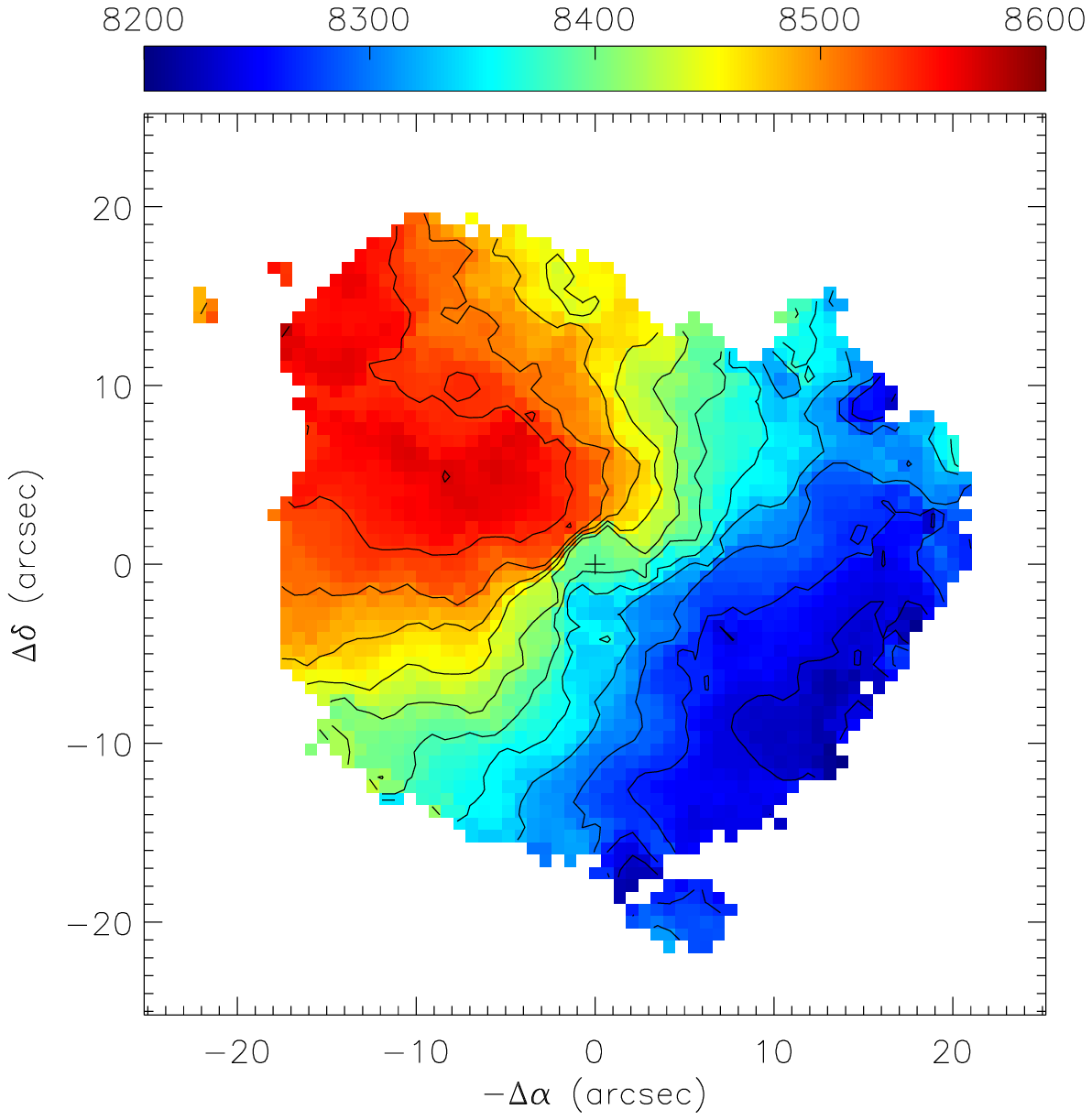}}
\hspace*{0cm} (b) \hspace{4cm}(c)\\
 \centerline{
\includegraphics[width=4cm]{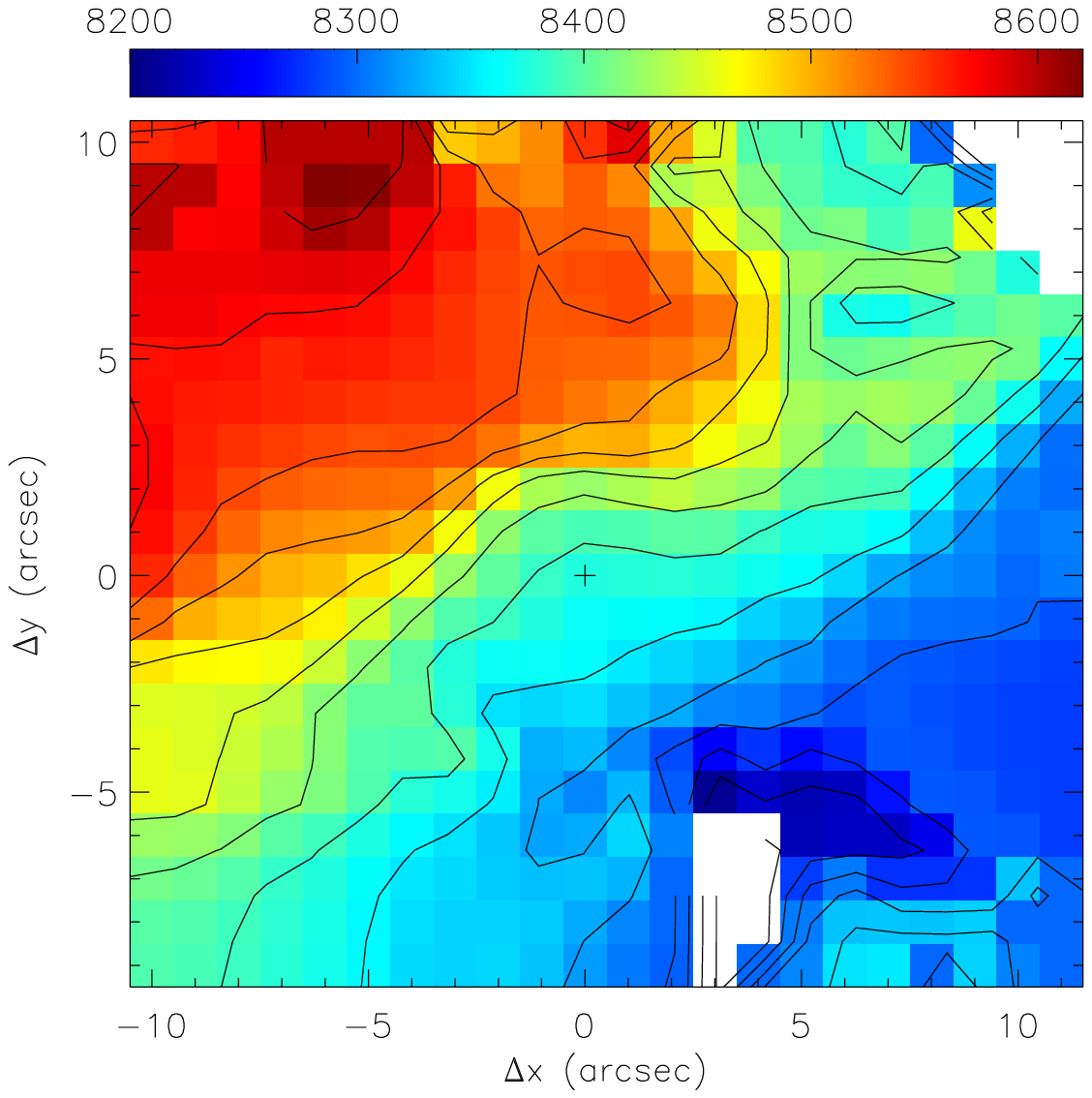}
\includegraphics[width=4cm]{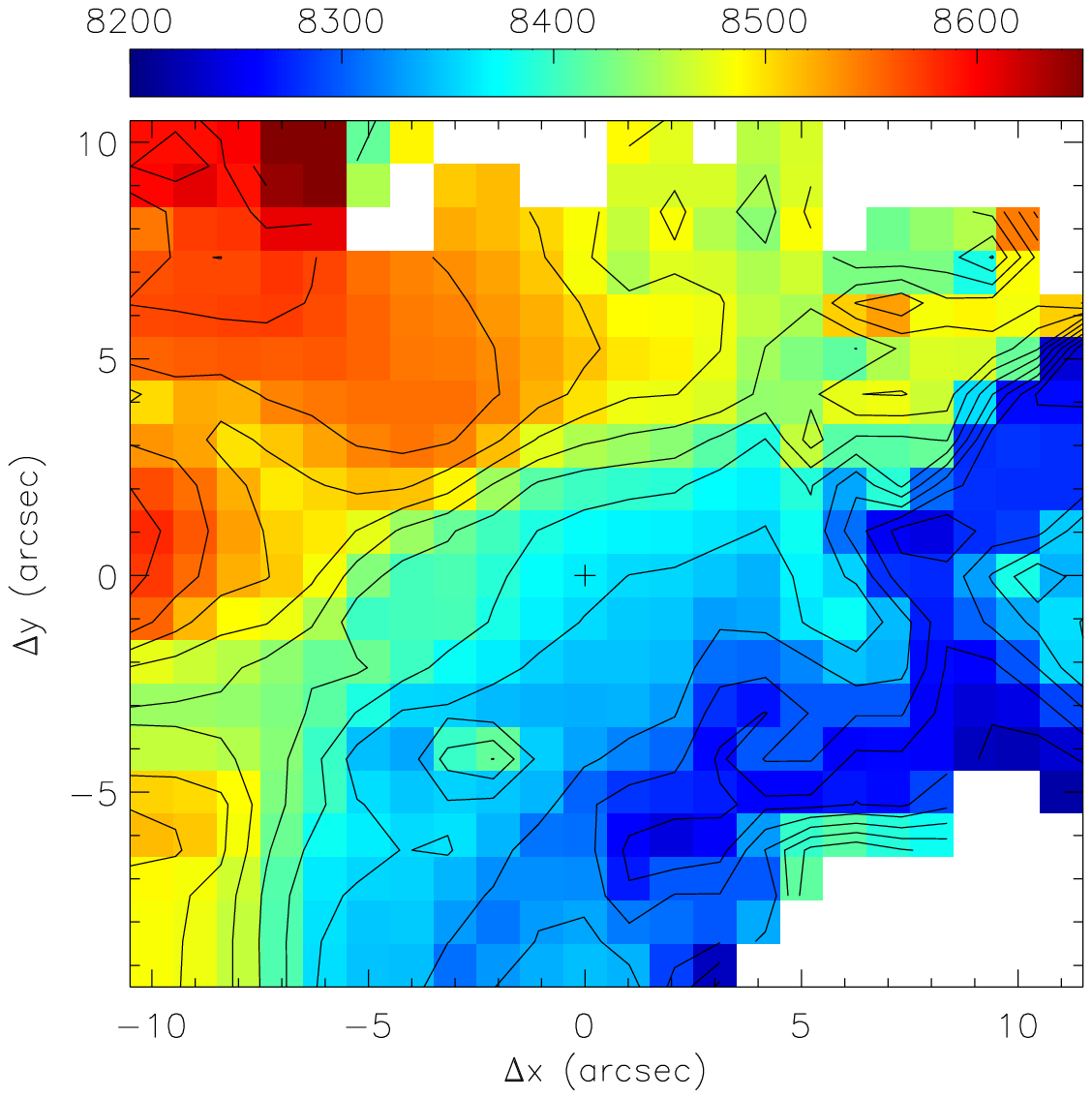} }
\hspace*{0cm} (d) \hspace{4cm}(e)\\
\centerline{
\includegraphics[width=4cm]{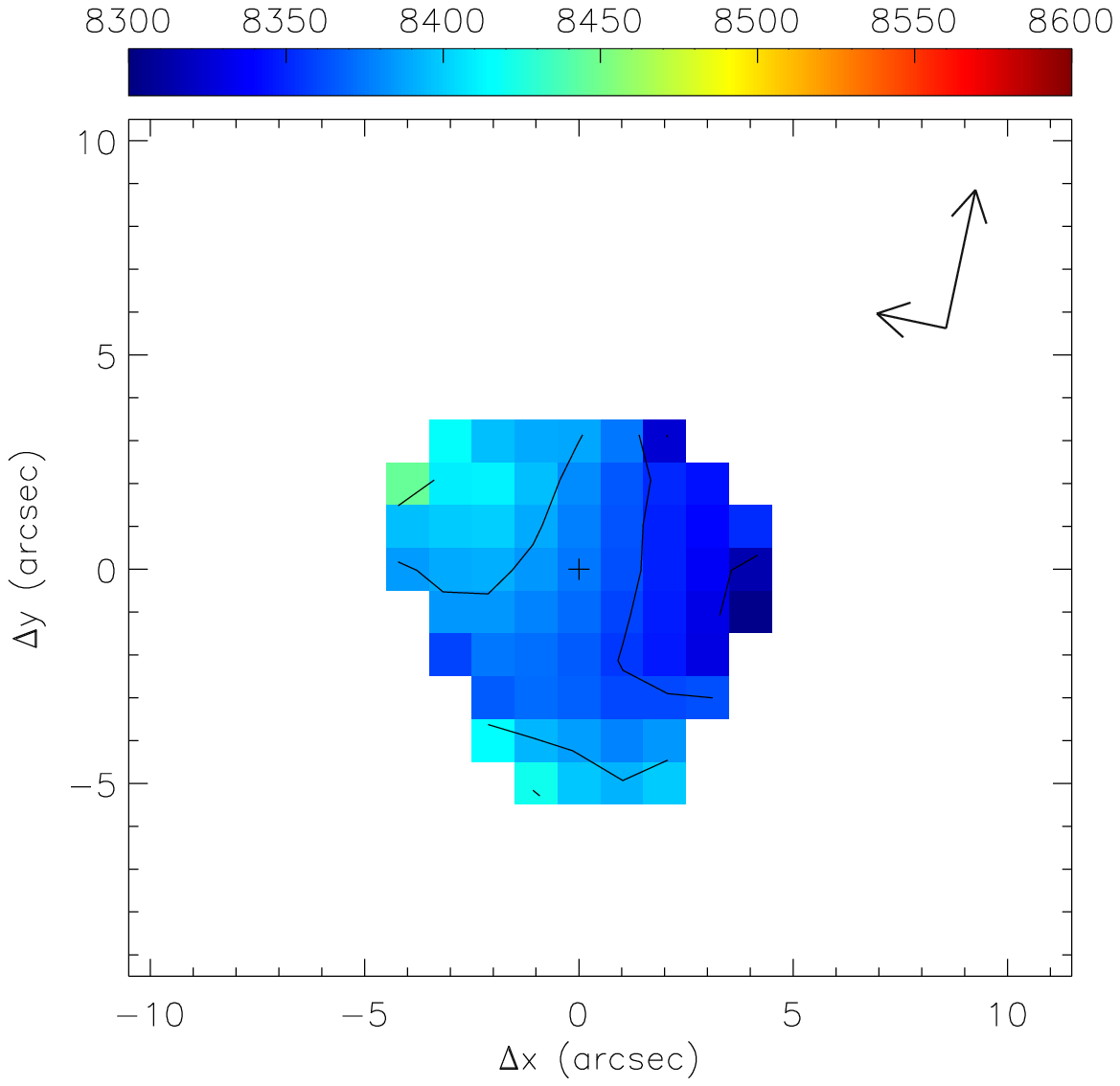}
\includegraphics[width=4cm]{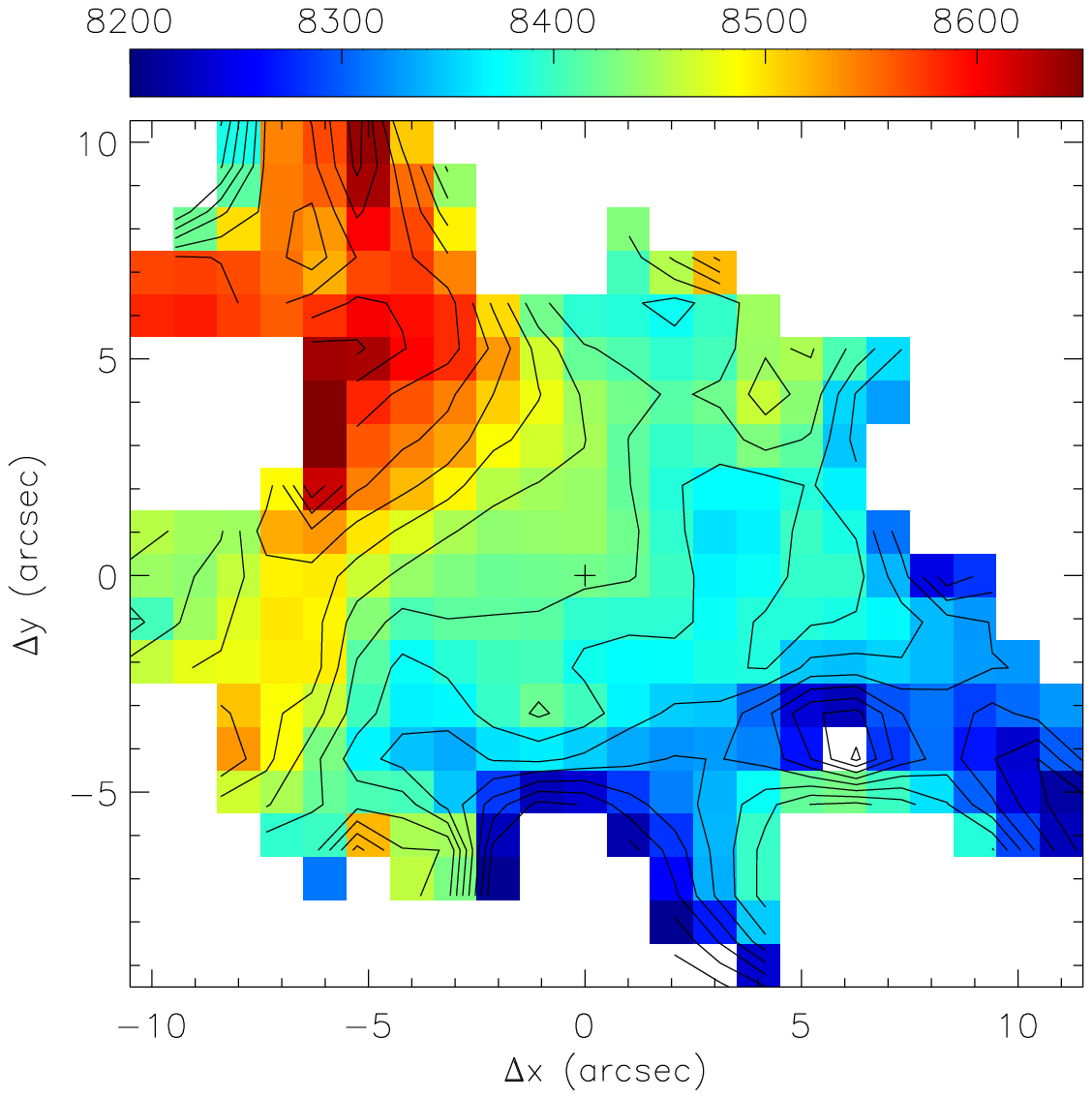}
}
 \caption{Radial velocity fields of NGC~6104 (in $\km$): (a) in \Ha emission line, constructed from FPI data; MPFS observational data in the
(b) \Ha, (c) [NII], (d) [OIII] emission lines and (e) the stellar
velocities. The isovelocities are everywhere drawn at intervals of
30 $\km$, the central contour  corresponds to the assumed systemic
velocity of $8418 \km$ . }
\end{figure}

\subsubsection{Long-slit spectroscopy.}

To study the galaxy's environment, we took long-slit spectra of
four faint galaxies, candidate dwarf companions of NGC~6104 (every
time the slit crossed two neighboring objects) and a spectrum of
the asymmetric feature in the image of the elliptical galaxy CGCG
196-022 close to it. The data were reduced using a software
package running in IDL. The reduction steps  were briefly
described by Afanasiev and Moiseev (2005).

\section{Surface photometry for NGC 6104.}

Figure 3 shows (a) the galactic image from the Digital Palomar
Observatory Sky Survey, (b) our deep \Rc-band image, and (c) the
WFPC2/HST image. The HST observational data have already been
briefly described by Malkan et al. (1998), who pointed out an
asymmetric ``perturbed'' shape of the tightly wound spiral arms
emerging from the bar. However, an examination of all the
available (including \Ha) images leads us to conclude that a
pseudo-ring is most likely present here, while the asymmetric
spirals are traceable in the outer regions at $r > 20''$.

Figure 4 shows the radial variations in the shape of elliptical
isophotes with radius $(r)$. The ellipticity peak at $r >10''$
corresponds to the end of the circumnuclear bar elongated along
the position angle $PA = 88^\circ$. As $r$ increases, the isophotal
ellipticity decreases sharply, while PA remains almost constant up
to $r \approx15''$. This behavior is attributable to the above
mentioned ring with a slight inner ellipticity that is noticeable
in all images and that has bluer colors than the rest of the
galactic disk. This feature caused by the presence of a younger
stellar population, the emission from HII regions ionized by O-B
stars is the most expressive indicator of ongoing violent star
formation in the ring. Thus, according to the FPI map, 70\% of all
galactic \Ha luminosity is concentrated in the ring (and only 5\%
in the AGN). The total observed \Ha luminosity of NGC~6104 is $2.5
\times 10^{41}\, erg\, s^{-1}$ or $1.3\times 10^{42}\, erg\,
s^{-1}$ after its correction for the internal extinction in the
galaxy\footnote{Here, we assumed that the internal extinction in
all starforming regions is the same, $A(H_\alpha)\approx 1.8^m$
(the reddening in \Ha that we obtained from the \Ha-to-\Hb flux
ratio using MPFS spectra for the HII region northeast of the
nucleus).}.  According to the formula from Kennicutt (1998), a
relatively high star formation rate of $SFR\approx 10\,M_\odot
yr^{-1}$, with $7\,M_\odot yr^{-1}$ for the ring alone,
corresponds to this luminosity.

The shape of the ring deviates from a circle even in the galactic
plane. This is clearly shown by Fig. 3c, which was constructed
according to the disk orientation obtained by analyzing the gas
velocity field (see below). We see that the ring remains
elliptical (with an axial ratio of $\sim0.85$); the boundary ring
radii were assumed to be $11''$  and $18''$. The southern half of
this structure forms an almost regular semiellipse, while its
northern half is more irregular (perturbed) in shape. The almost
rectilinear fragment of the spiral arm between the bar and the
inner edge of the ring looks particularly strange. Also, the ring
center formally determined from isophotes is displaced by several
arcseconds to the southeast of the nucleus. The displacement of
the center of the star formation ring is also immediately apparent
in the \Ha image (Fig. 2a). Below, we will return to the
discussion of peculiarities in the shape of this structure.

The outer shells firstly found by us in the galaxy are its most
intriguing morphological features. These shells are completely
invisible in the images from the Digital Palomar Observatory Sky
Survey (Fig. 3a) and are barely detected in the images from the
Sloan Digital Sky Survey (SDSS). However, they are clearly
distinguishable in our images in all three color bands at a
surface brightness level of $24.5-25.5^m/\Box''$. These shells are
most conspicuous southwest of the galactic center at distances of
more than $50''$  (28 kpc) from the nucleus, i.e., in the opposite
direction and considerably farther than the northeastern feature
in the surface brightness profile (Fig. 3b). Many authors who
classified the galaxy as a peculiar one have already called
attention to this feature (see the Introduction for references).
Interestingly, despite the presence of this asymmetric feature in
the images, the mean brightness profile (averaged in elliptical
apertures elongated in the agreement with the galactic disk
orientation) is satisfactorily described by a classical
exponential law. When the radial brightness profiles were
examined, it was found that the spatial resolution of the images
from the 6-m telescope ($\sim$1\farcs5) was too low to reliably
separate the contributions from the compact bulge and the starlike
nucleus. Therefore, we used an image through the F606W filter from
the HST archive for a detailed analysis.

\begin{figure}
(a)\\
\centerline{\includegraphics[width=8.5cm]{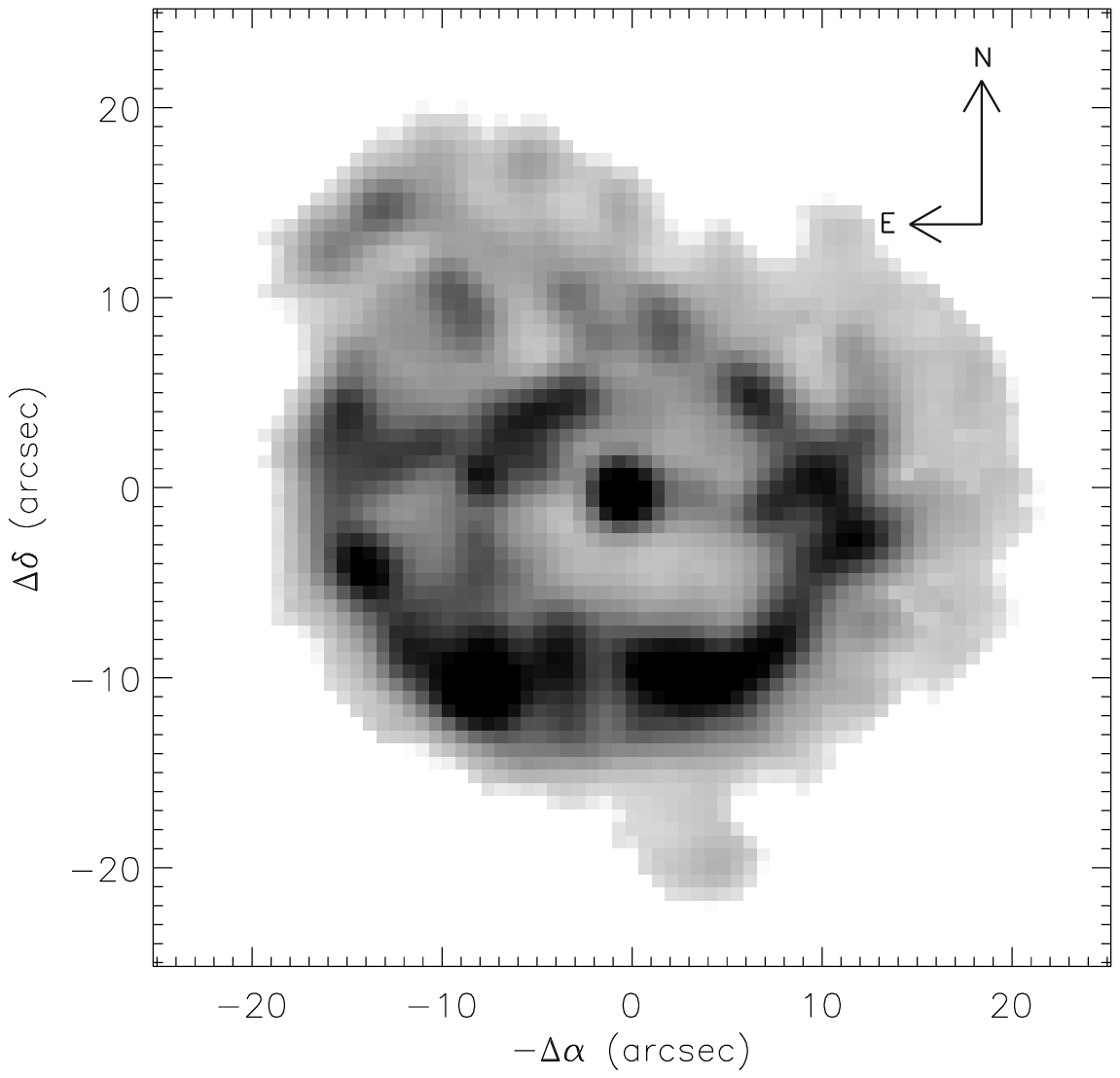}}
 (b) \hspace{4cm}(c)\\
\centerline{
\includegraphics[width=4cm]{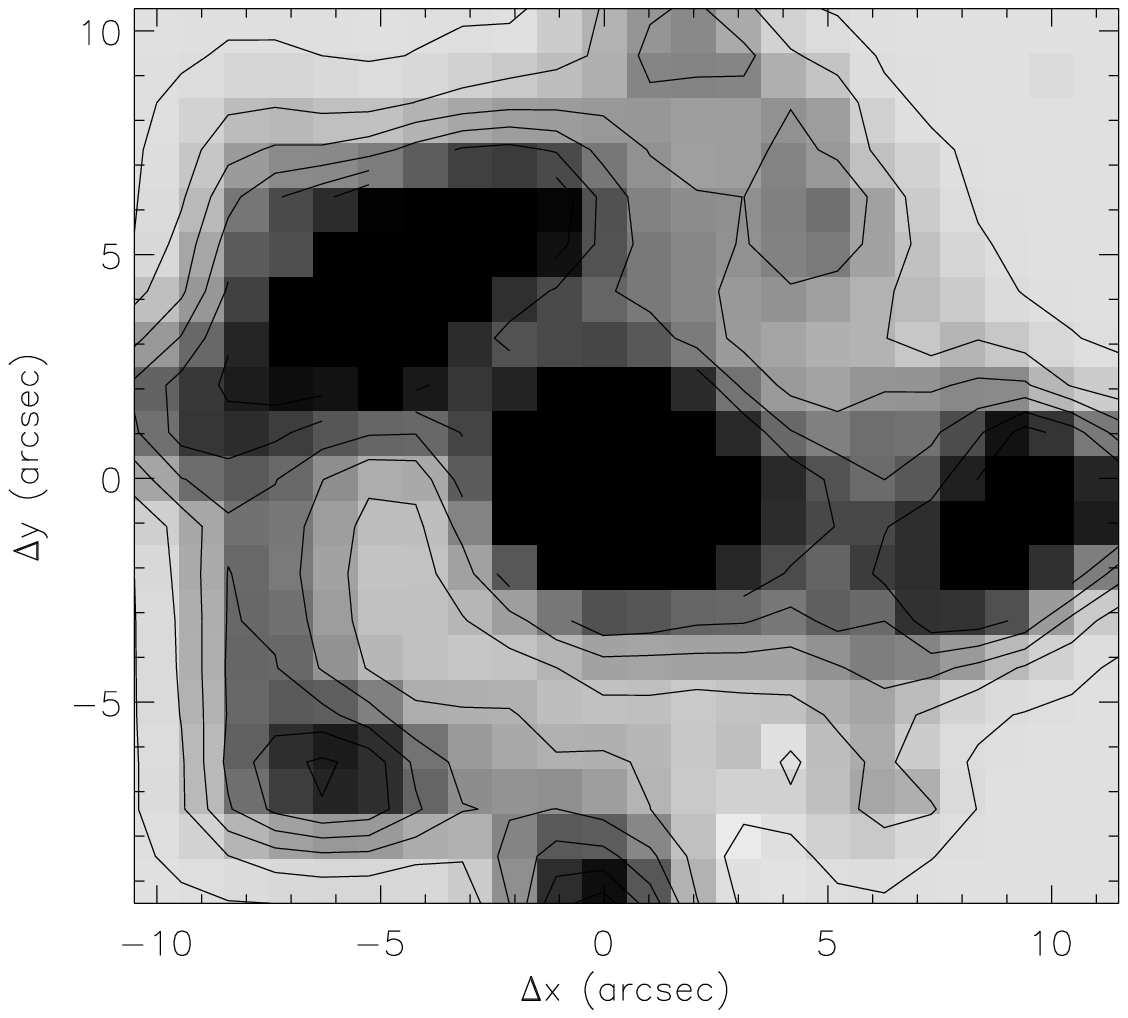}
\includegraphics[width=4cm]{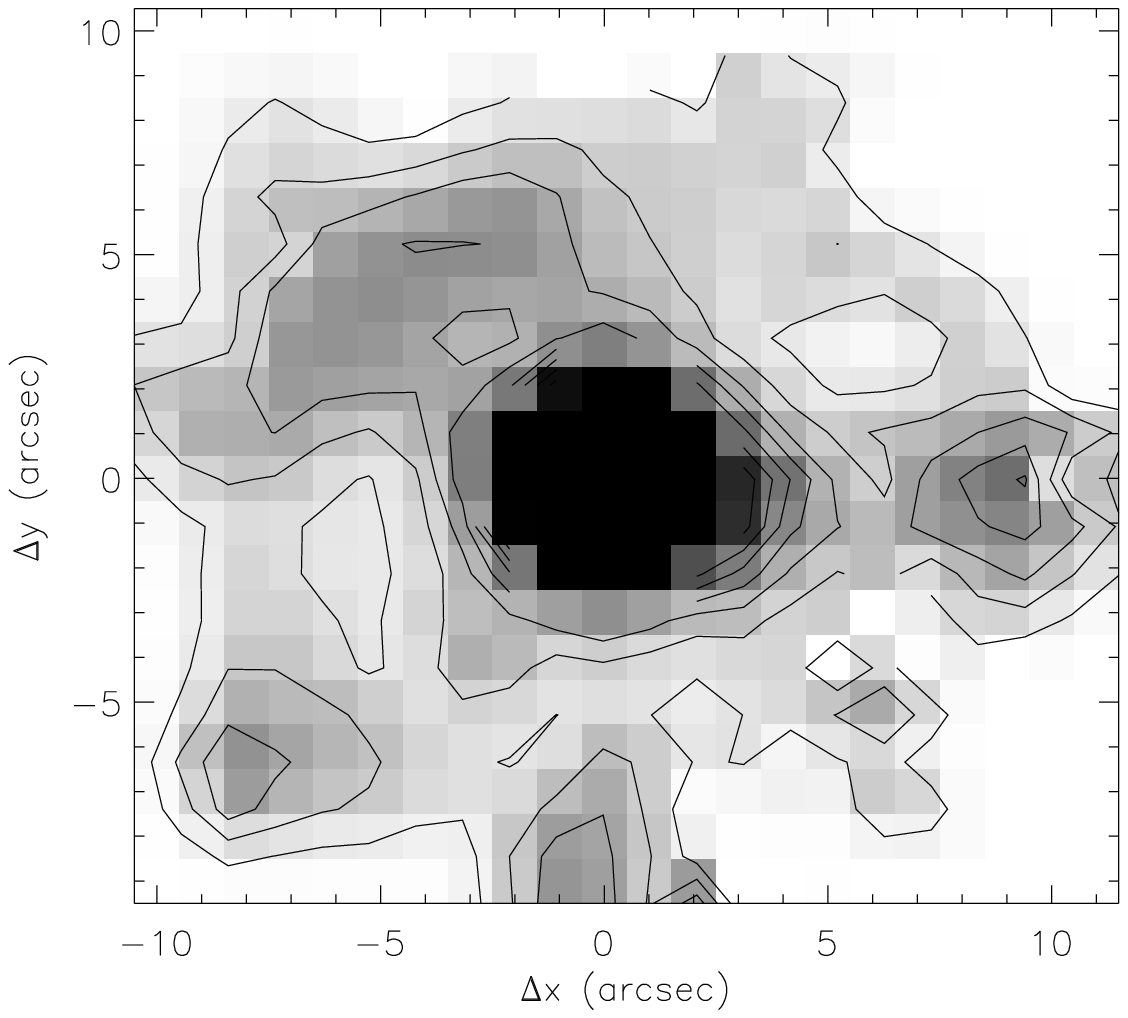}
}
 (d) \hspace{4cm}(e)\\
\centerline{
\includegraphics[width=4cm]{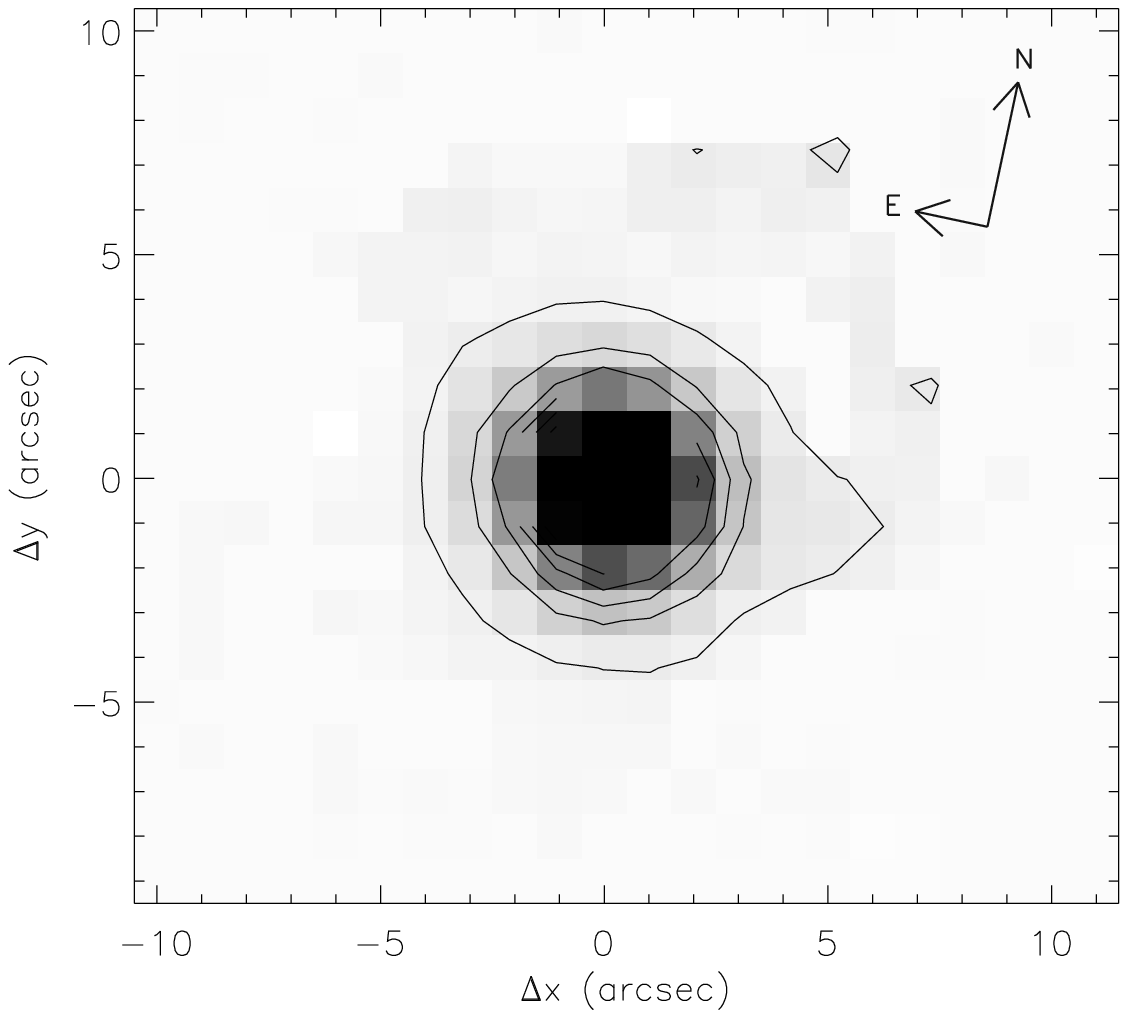}
\includegraphics[width=4cm]{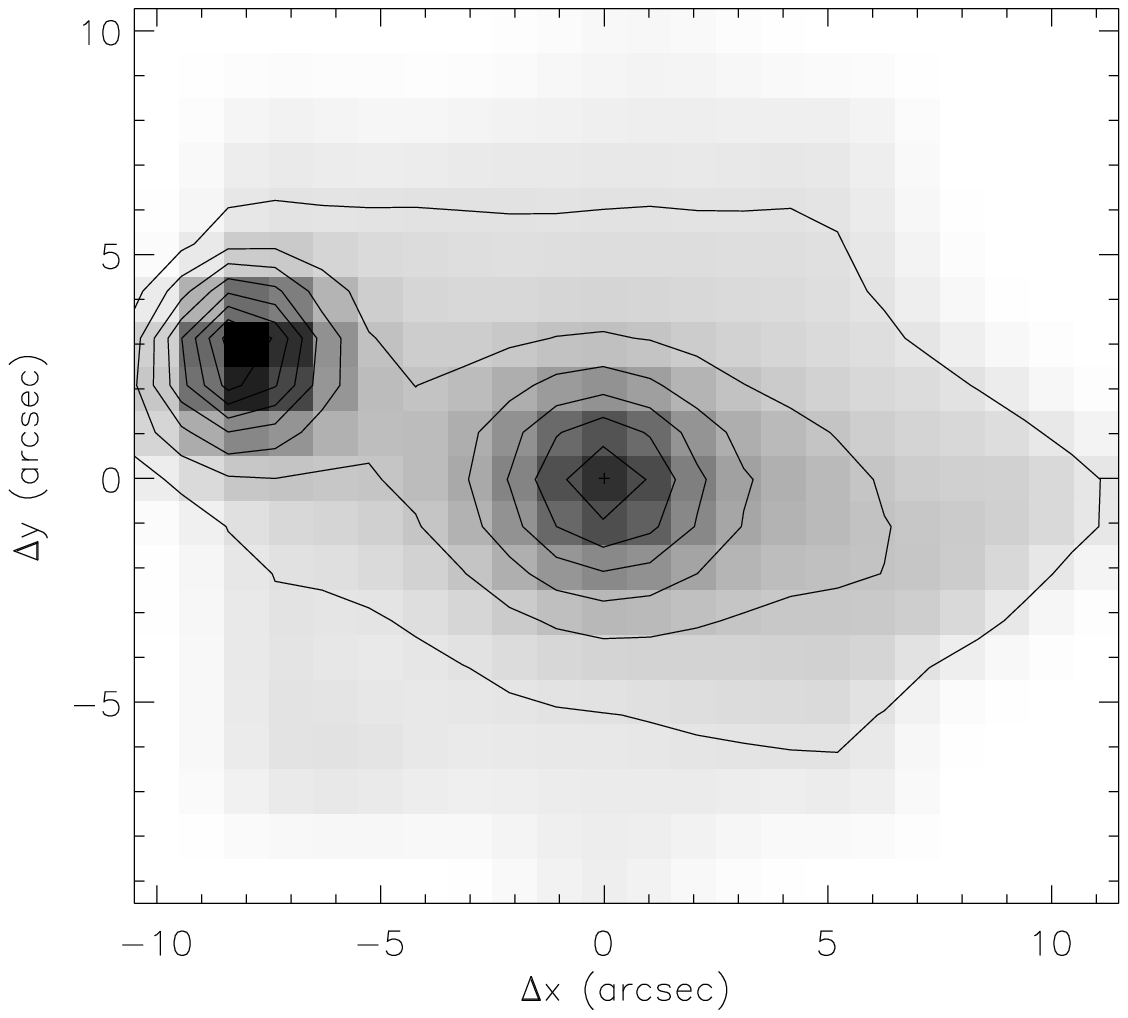}
} \caption{Monochromatic images of NGC~6104: (a) \Ha (FPI) map.
MPFS maps: in the (b) \Ha, (c) [NII], (d) [OIII]  lines and (e)
in the continuum near $\lambda6600$\AA.}
\end{figure}

We decomposed the surface brightness profile into standard
components: an exponential disk (the radial scale length
$r_{disk} = $9\farcs4 coincided with the scale determined from the
deep \Rc-band image), a de Vaucouleurs bulge (with an effective
radius of $r_e = $1\farcs7), and a starlike nucleus. The derived
component luminosity ratio $L_{bulge}/L_{disk} =0.14$ corresponds
to the Sc morphological type (according to Simien and de
Vaucouleurs, 1986).

After the deprojection of the images for NGC~6104 to the face-on
position and the subtraction of the two dimensional dusk+bulge
model constructed from the surface brightness profile
decomposition, it became clear that the outer shells, which are
clearly seen in the images in all three color bands, are a single
structure that bends around the disk of NGC~6104 (Fig.3d). This is
most closely resembles a companion galaxy that has been disrupted
almost totally by tidal forces. Similar structures that are formed
when galaxies are disrupted by tidal forces have been described in
many papers, as well observational ones as devoted to numerical
simulations (e.g., Jogee et al. 1999). In the case under
consideration, we most likely observe the merging of a companion
with NGC~6104.

According to our estimates, the total luminosity of the outer
low-surface-brightness filament accounts for $\sim2\%$ of the
galactic total R-band luminosity (minus the luminosity from the
starlike AGN). The fractional luminosity increases to 3-7\% if the
bright fragments located to the northwest and the northeast of the
nucleus at $r = 20''-30''$ are assumed to also belong to this
tidal structure (Fig. 3d). Thus, there can only be merging exactly
with a dwarf companion (with $M_R > -17^m$) whose mass accounts
for no more than 2-7\% of the mass of NGC~6104.

\section{Spectrophotometry and ionization diagrams.}

The galactic nucleus, bar and a bright field star are
distinguished in the MPFS continuum images. The bar and the star
formation ring are clearly seen on the maps in the most intense
emission lines (except [OIII] $\lambda5007$). In the [OIII] image,
only the bright nucleus is seen and there are no features far from
the center.

To determine the ionization sources of the central galactic
regions, we constructed ionization diagnostic diagrams from MPFS
data. Using the ratios of lines with different excitation
mechanisms, we can determine the regions where thermal (young hot
stars), non-thermal (AGN), or shock ionization dominates. The
boundaries that separate the regions with different ionization
sources (AGN, HI or LINER) in Figs. 5a and 5b were taken from
Veilleux (2002). We distinguished three regions in the MPFS field
of view: a nucleus, a bright region at the end of the bar in the
Ha, [NII] and [SII] lines and points belongs to the bar at
distances of $3''-7''$ from the nucleus. In both our diagrams, the
points belonging to the nucleus lie in the region of ionization by
nonthermal radiation. However, according to the [OIII]/\Hb flux
ratio criterion, about one third of all points from the nuclear
region lie near the AGN--LINER boundary, i.e., the gas can also be
partially ionized here by shocks. In the next section, we will
return to this issue in the discussion about nuclear jet.

All points from the region of the northeastern knot fall into the
photoionization region; on our \Ha map (Fig. 2a), one of the
brightest HII regions at the end of the bar corresponds to these
points. In both diagrams, the points from the bar region are
located in the transition region between AGN ionization  and shock
ionization; several points in the [SII]/\Ha diagram fall into the
region of photoionization by young hot stars. Ionization by
shocks, which can be associated both with the edges of the bar
(where dust lanes are noticeable in the HST image) and with the
nuclear jet (which, however, may be not so large, see below), is
apparently also added to the two main competing mechanisms
(photoionization/non-thermal source) at the points belongs to the
bar region and the AGN.

\section{Analysis of the kinematics of ionized gas and stars.}

We determined the orientation parameters of the galactic gaseous
disk using the model of pure circular rotation of a thin disk in
which the inclination to the line of sight, $i_{0}$, and line of
nodes position angle, $PA_0$, did not change with radius. Here, we
analyzed the  large-scale \Ha velocity field (excluded the points
at $r < 9''$, where significant peculiar motions take place). The
following values were obtained by $\chi^2$ minimization of the
observed velocities deviations from the model approximation: $PA_0
= 56 \pm 3^\circ$ and $i_0 = 39 \pm 5^\circ$; these values are in
a good agreement with the data of the orientation of the outer
disk isophotes (Fig. 4).
\begin{figure*}
(a) \hspace{9cm} (b)\\
\includegraphics[width=8.5cm]{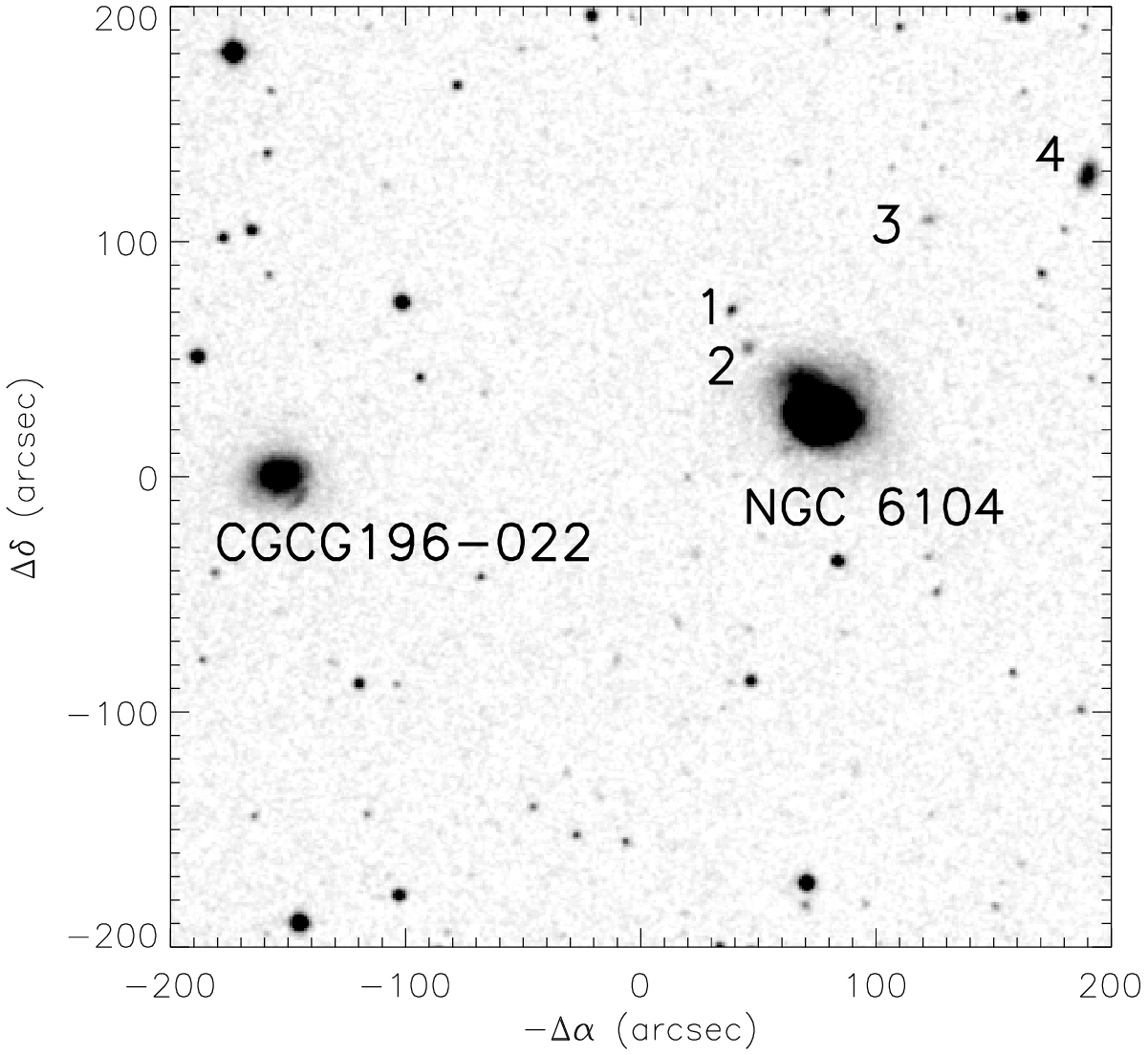}
\includegraphics[width=8.5cm]{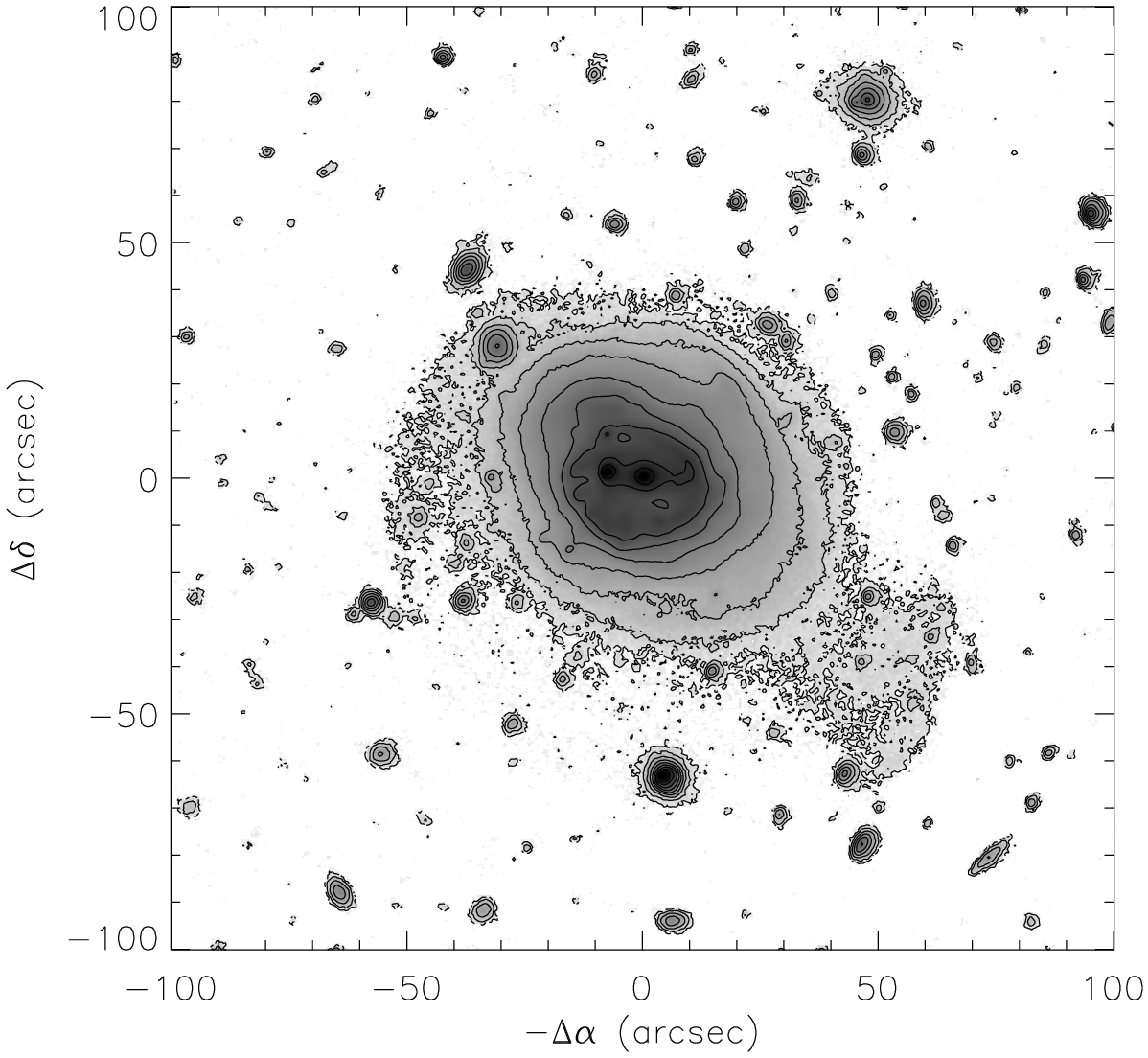}\\
 (c) \hspace{9cm} (d)\\
\includegraphics[width=8.5cm]{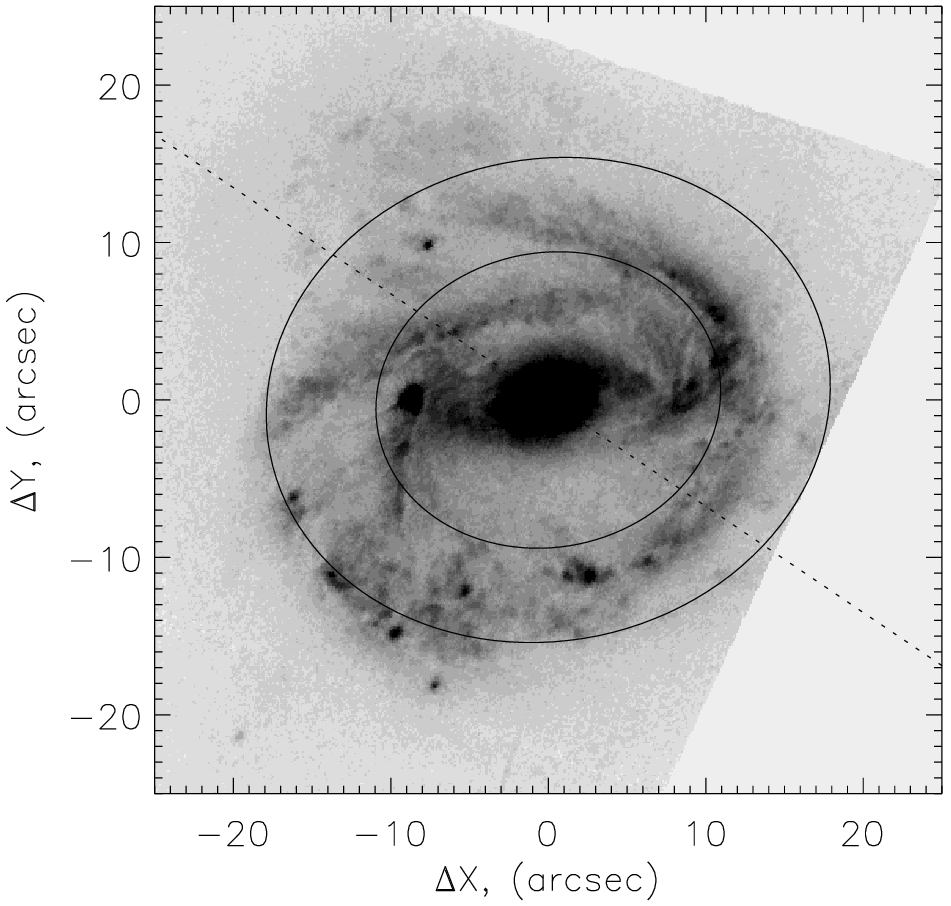}
\includegraphics[width=8.5cm]{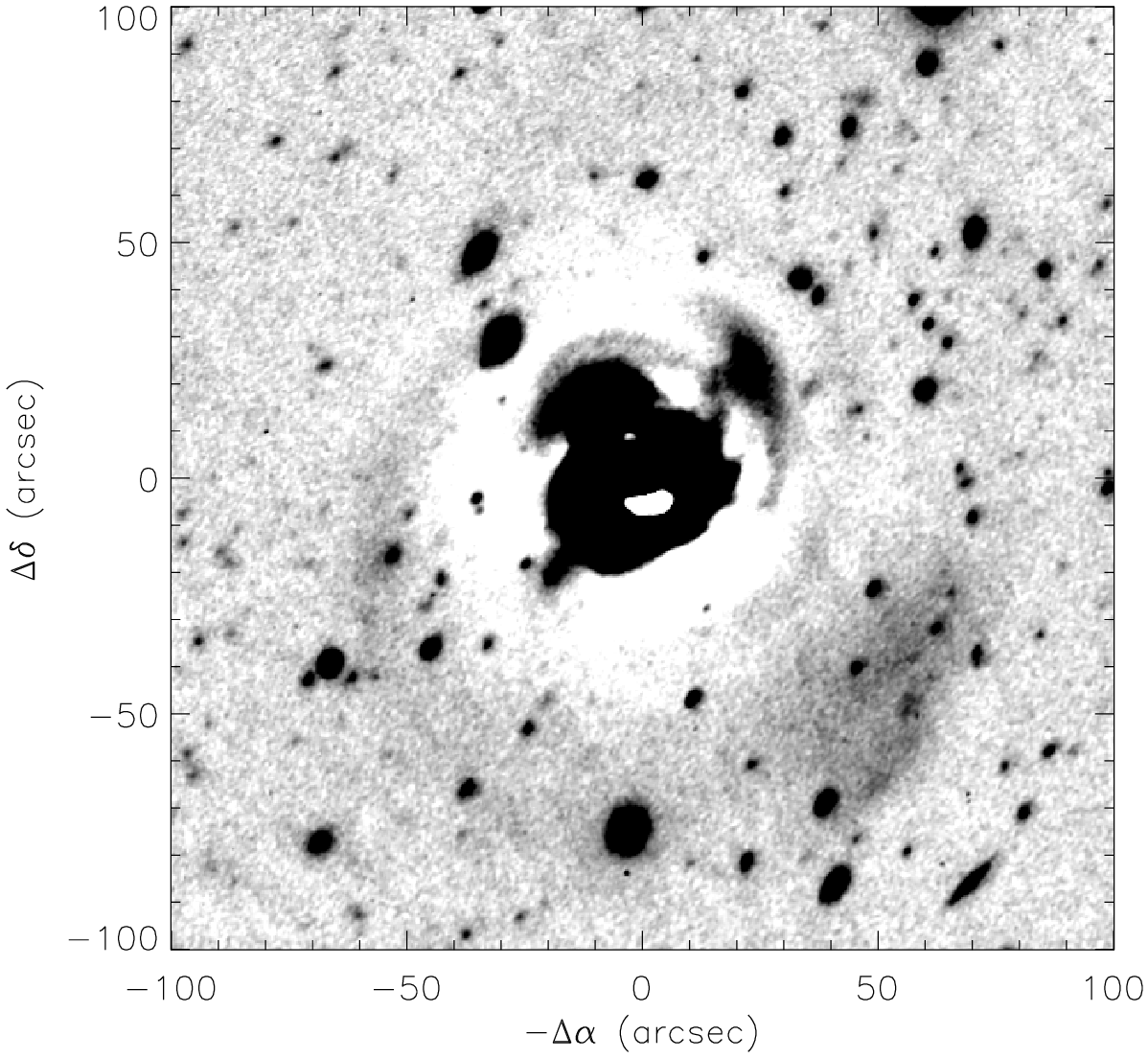}
\caption{(a) POSS2 image of the NGC~6104 field, the numbers mark
four candidates for dwarf companions. (b) \Rc image of NGC~6104 in
the magnitude scale, the outer isophote corresponds to a surface
brightness of $25^m/\Box''$, the isophotal step is $1^m$. (c) HST
image with the F606W filter deprojected to the face-on position.
The dotted line corresponds to the line of nodes, the ellipses
mark the adopted maximum and minimum radii of the ring with
allowance made for its ellipticity. (d) Residual brightnesses of
the \Rc-band image for NGC~6104 after deprojection to the face-on
position  and the subtraction of the disk model.}
\end{figure*}

\begin{figure}
\includegraphics[width=8.8 cm]{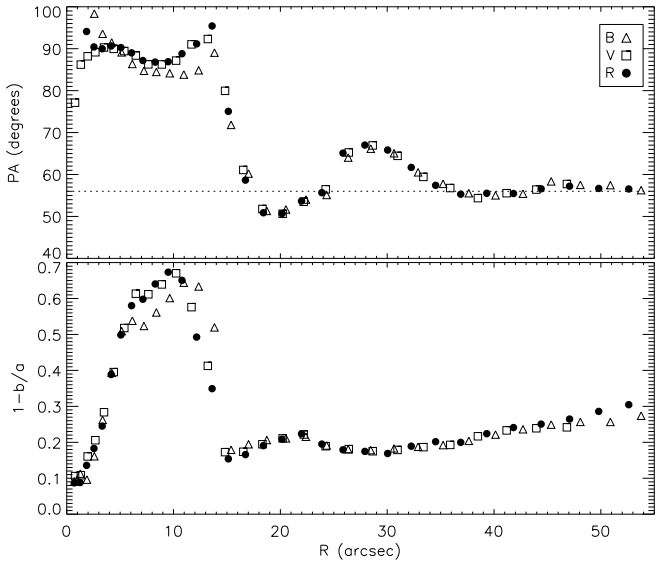}
\caption{Results of the isophotal analysis of the images for
NGC~6104 in three color bands: radial variations in the position
angle of the major axis of the isophotes (a)  and their
ellipticity (b). The dotted line marks the location of the line of
nodes.}
\end{figure}

Then, we analyzed the noncircular motions by the tilted-ring
method. The velocity fields were broken down into elliptical
$1''$-wide rings oriented according to the values of $PA_0$ and
$i_0$ obtained above. In each ring we determined the optimal
position angle of the kinematic axis $PA_{kin}$, the mean rotation
velocity $V_{rot}$ and the systemic velocity $V_{sys}$ in the
approximation of circular rotation. See,   Moiseev et al. (2004)
for details of the method and references to original papers. The
center of the rings was fixed at the galactic nucleus. The
symmetry center of the velocity field is slightly (by 1\farcs5)
displaced relative to the isophotal center of the continuum image,
we attribute this displacement to the effect of the noncircular
motions that we detected. The constructed dependences
$PA_{kin}$(r) and $V_{rot}(r)$ are shown in Figs. 6a and 6b. The
velocity fields of ionized gas and stars are in a good agreement
with one another, demonstrating a similar turn $PA_{kin}$ relative
to the line of nodes ($PA_{0}$) in the bar region (Fig. 6a). The
relatively large scatter of MPFS data points is a result of a
lower MPFS measurement accuracy than the accuracy of kinematic
estimates in \Ha with FPI, which has a higher spectral resolution
and a smaller pixel size. However, there are also differences in
the kinematics of gas and stars.

Firstly, Fig. 6b shows that the rotation amplitude for the stellar
component at $r = 3''-8''$ is appreciably (by $\sim50 \km$)
smaller than for the ionized gas in \Ha. This mismatch results
from a considerably higher stellar velocity dispersion than the
gas one (the asymmetric drift effect widely known in stellar
dynamics). More unusually, about half of all velocity measurements
in the [NII] doublet are also indicative of a systematically lower
rotation velocity than that in \Ha. In our opinion, this is
attributable to a gas deceleration at the shock front that emerge
at the bar edges where the fraction of the gas emission in
forbidden lines, including that in [NII], increases. This effect
was considered by Moiseev (2002b) for several barred galaxies.
Recall that, according to the MPFS diagnostic diagrams, the
emission from some regions inside the bar just points  to the
shock ionization mechanism.

Secondly, an excess of ``blue'' ionized gas radial velocities,
which is particularly clearly seen in Fig. 1a, is observed in the
nucleus. As a result, the formally measured $V_{sys}$ at $r < 3''$
in the ionized gas velocity fields are appreciably lower than the
mean systemic velocity at large radii ($V_{sys} = 8418 \km$),
while in the stellar component $V_{sys}(r)\approx const$. We
interpret this excess of negative velocities as a signature of
radial gas motions toward the observer; the outflow source itself
is not resolved (i.e., its angular size is less than $1.5-2''$).
We observed similar signs of nuclear outflow when analyzed
integral-field spectroscopy for the galaxies Mrk 315 (Ciroi et al.
2005) and NGC2273 (Moiseev et al. 2004). Note that in all these
objects, we observe not the high-velocity jet itself emerging from
the AGN, but the result of the interaction between the radio jet
and interstellar clouds, the formation of an expanding cocoon of
hot gas. One of the most characteristic examples was considered by
Capetti et al. (1999), who observed Mrk 3. The size of such a
structure (radio jet+cocoon) in Seyfert galaxies is only several
hundred parsecs and it is not surprising that our spatial
resolution is not enough to study it in NGC~6104 in detail. In NGC
6104, our measurements give an estimation of the gas outflow
velocity along the line of sight $\sim30-40 \km$ in the \Ha,
[NII], [SII] and [OIII] lines. In the latter case, the formal
application of the tilted-ring model yields $PA_{kin} \sim
80^{\circ}$, which differs fundamentally from the measurements in
the remaining lines. Here, most of the [OIII] emission apparently
originates from the circumnuclear jet and it is inappropriate to
use the model of circular motion.

\begin{figure*}
(a)\hspace{5.5 cm}(b)\hspace{5.5 cm}(c)\\
\includegraphics[width=17cm]{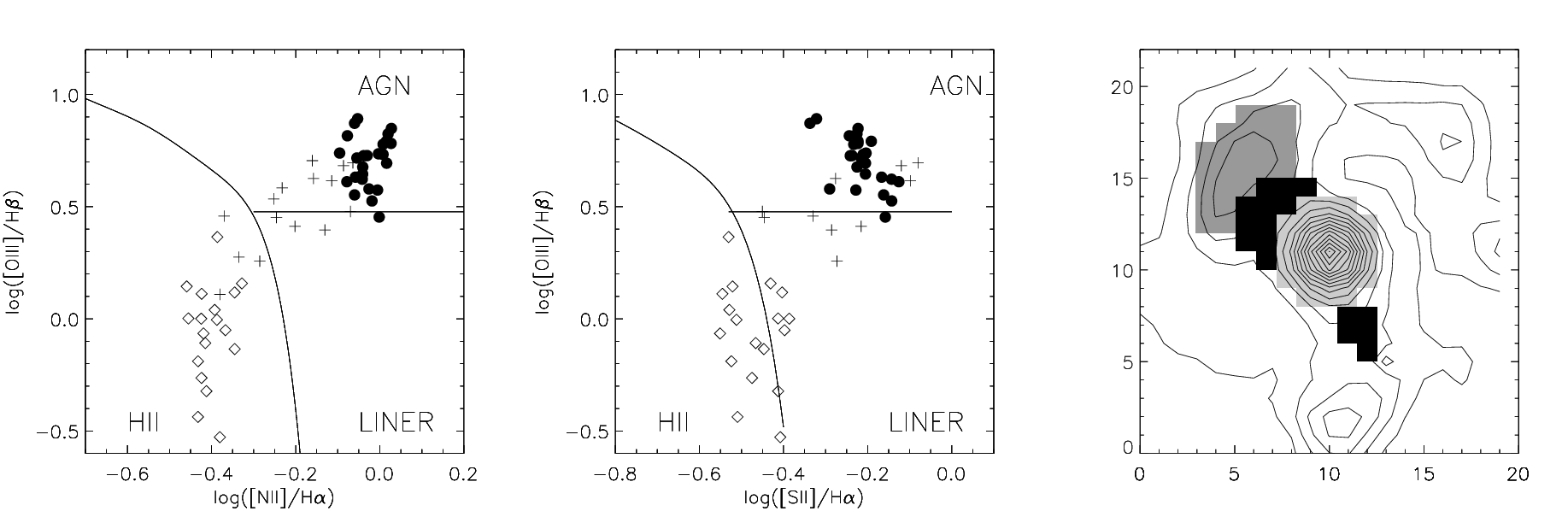}
\caption{(a, b) Diagnostic diagrams for NGC~6104. The filled
circles, diamonds, and crosses, correspond to the nucleus, the
star-forming region in \Ha, and the bar region,  respectively. (c)
The mask for separating the above-mentioned regions with \Ha
isophotes superimposed on it.}
\end{figure*}

It is possible that the points corresponding to the
nucleus in the ionization diagrams (Figs. 5a and 5b)
lie near the AGN/LINER boundary owing to the
interaction of the jet with the interstellar medium. It
should be noted that we give only a lower limit for
the outflow velocity, because the spectral resolution
is too low for the component associated with the gas
surrounding the jet to be directly identified; we detect only
the shift of the line profile as a whole. Finally, note that
we have failed to find any references in the literature
to radio observations of NGC~6104 with an angular
resolution high enough to detect this radio jet.

$PA_{kin}$ within $10''$  of the center turns in the direction
opposite to the line of nodes compared to $PA$ of the bar
isophotes (Fig. 6a). Such behavior of the position angle is in
agreement with both available observational data on the kinematics
of barred galaxies and with numerous model calculations of radial
motions toward the center in the region of a triaxial bar
potential (for references to original papers, see Moiseev et al.
2004). This turn of $PA_{kin}$ is better seen in the FPI data. Because of
the above-mentioned radial outflow from the nucleus, we should be
very guardedly in the estimates of $PA_{kin}$ at $r = 1''-3''$
based on such a simple rotation model, since the velocities of two
spatially separated components, the gaseous disk and the outflow
from the AGN, are projected onto the line of sight.

Significant radial (inflow) motions take place only inside
the bar, since the kinematic axis coincides with the
line of nodes of the disk even at $r \geq 10''$ (i.e., in the
ring region). We tried to estimate the amplitude of
the radial motions using a simple model that assumed
each point at a given radius $r$ to follow circular
rotation ($V_\varphi$) and to be involved in radial motions
($V_R$), so the observed radial velocity is the sum of the
components of both vectors:
\begin{equation}
V_{obs}(r) = V_{sys} + V_R(r)\sin\varphi\sin i + V_\varphi(r) cos\varphi \sin i.
\end{equation}
Here, the azimuthal angle in the galactic plane $\varphi$ is
measured from the line of nodes. We obtained $V_R =50\pm10 \km$ at
$r = 2''-8''$ and $V_R\approx 0$ at $r > 10''$,
in agreement with the previous conclusion.

Since few direct observational measurements for inflow motions in
bars are founded in the literature (they are considerably fewer
than the bar pattern speed estimates), each such measurement is of
considerable interest in discussing the role of bars in AGNs
fueling and secular evolution of galaxies. Our value ($\sim50
\km$) is close to that obtained by Jogee et al. (1999) for the
nuclear bar of NGC~2782 ($15-45 \km$), but is considerably lower
than the velocity of inflow motions in the much higher-contrast
bar such as in NGC1530 ($85-200 \km$; Regan et al. 1997).

\section{Origin of the ring structure.}

As we have already noted above, a ring $\sim6$ kpc in radius that
differs from the remaining galactic regions by a high star
formation rate is clearly seen in all optical images of NGC~6104.
The following scenarios of the formation of this structure are
possible.

The first, most obvious cause for a barred galaxy is the formation
of a ring at one of the dynamical resonances. Comparison of the
results of numerical simulations with real ring galaxies (for
references, see Buta and Combes 1996) shows that the rings are
formed near the inner and outer Lindblad resonances (the nuclear
and outer rings, respectively) and at the ultraharmonic (1:4)
resonance near the end of the bar (the ``inner ring''). The size
and location of the ring in NGC~6104 exactly corresponds to the
latter case.

The second cause is considerably rarer, but possible in an
interacting system is a collisional ring, i.e., an expanding ring
density wave generated by the passage of a massive companion
through the galactic disk. However, we have found no ring
expansion in NGC~6104; significant non-circular motions (different
from an expansion) are observed only in the inner region associated
with the bar. Similarly, for obvious reasons that follow from the
galactic morphology and kinematics, a polar ring is not possible
either.

A further argument for the resonant nature of the ring is its
elongated shape with an axial ratio of 0.85 in the galactic plane.
The ring is elongated in the same direction as the bar, which must
be in the case for rings at resonances within the corotation
radius, in agreement with observations (Buta 1986) and model
calculations (Schwarz 1984). Therefore, it would be interesting to
independently determine the positions of the main resonances in
the galaxy by measuring the bar pattern speed $\Omega_p$.

Numerous methods for estimating $\Omega_p$ are described in the
literature (see Hernandez et al. 2005), but the overwhelming
majority of them are indirect ones based on additional assumptions
about the bar formation mechanisms or on numerical simulations.
However, in recent years, a model-independent method based on
considerably more general postulates formulated twenty years ago
by Tremaine and Weinberg (1984) has come into wide use. Let a
rotating (gaseous or stellar) medium distributed in a flat disk,
whose motion is characterized by a single pattern speed $\Omega_p$
and for which the continuity equation holds, in any case, on time
scales of several disk rotations. The observed parameters at each
point in the disk are its surface brightness and line-of-sight
velocity. Then,
\begin{equation}
\langle V\rangle=\langle X \rangle\Omega_p \sin i,
\end{equation}
 where the $x$ -axis is parallel to the galaxy line of nodes, $\langle V
\rangle$ is the weighted mean observed radial velocity in a
narrow strip along this axis, and $\langle X \rangle$ is the
weighted mean coordinate along this strip; the surface brightness
is used as a weight factor. An additional condition is a linear
relation between the surface brightness and surface density of
the medium under study; the latter must become zero at the disk
edges. The Tremaine-Weinberg (TW) method implies an infinitely
thin, flat disk and ignores any vertical motions. Note that,
despite the traces of interaction in the galactic outer regions,
the inner gaseous disk of NGC~6104 may be considered flat, at
least where there are points in the FPI velocity field. Indeed,
as follows from Fig. 6a, $PA_{kin}$ is almost equal to the mean
value at $r = 10''-25''$; moreover, there are no significant
variations in the parameters of elliptical isophotes at larger
distances, $r = 35''-55''$ (19-30 kpc) (Fig. 4). Therefore, the
warp of the disk is possible only in the outermost region.

\begin{figure*}
(a) \hspace{8.3cm} (b)\\
\includegraphics[width=8cm]{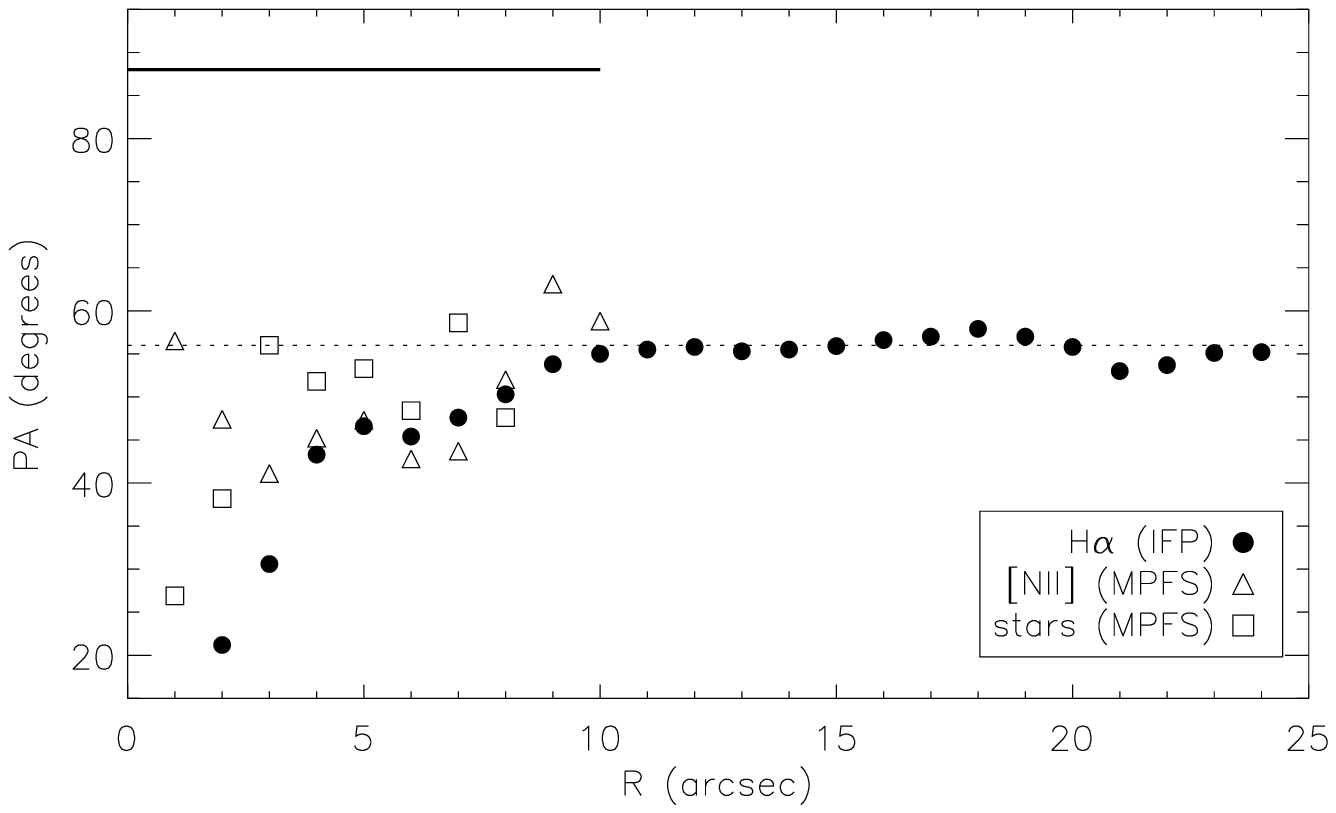}\hspace{0.5 cm}
\includegraphics[width=8cm]{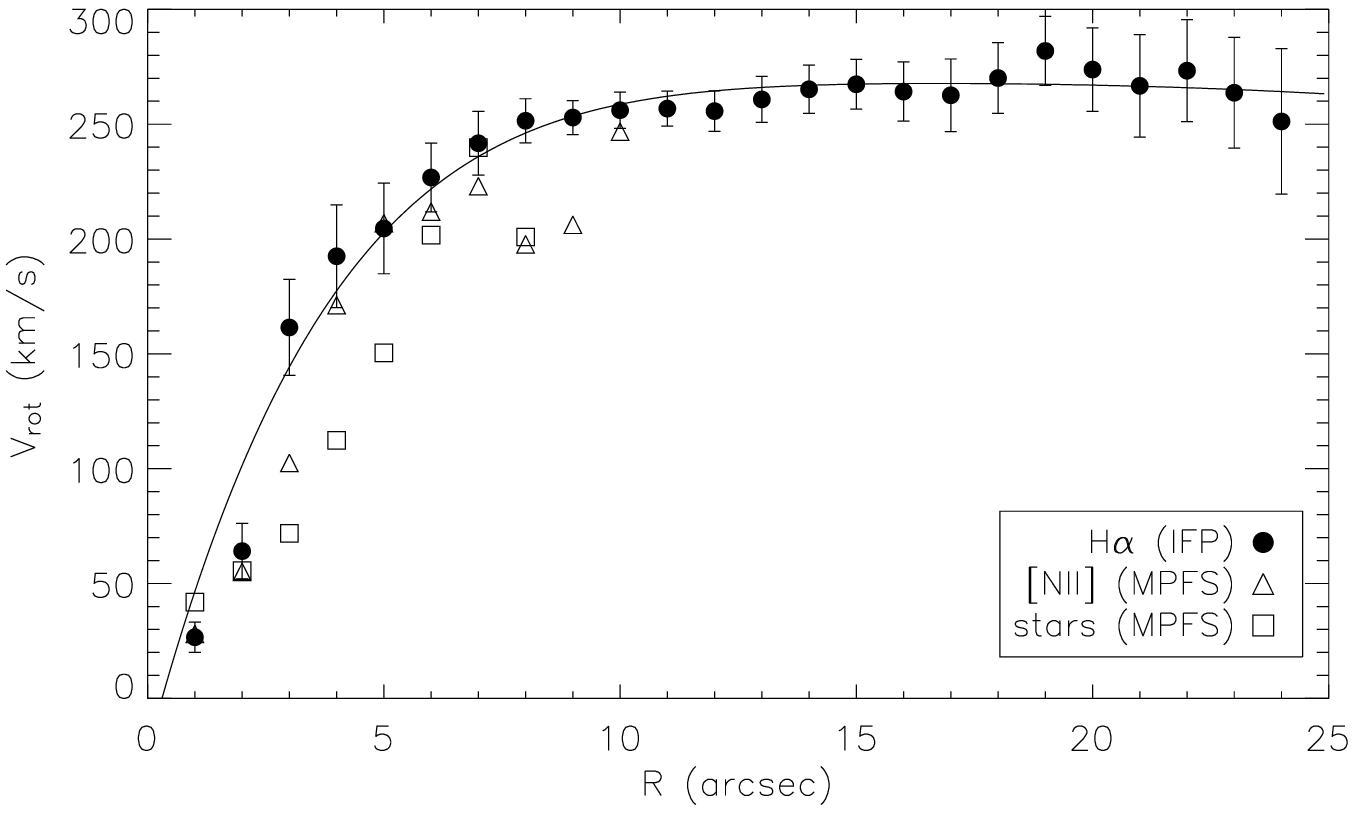}
\\
(c) \hspace{8.3cm} (d) \\
\includegraphics[width=8cm]{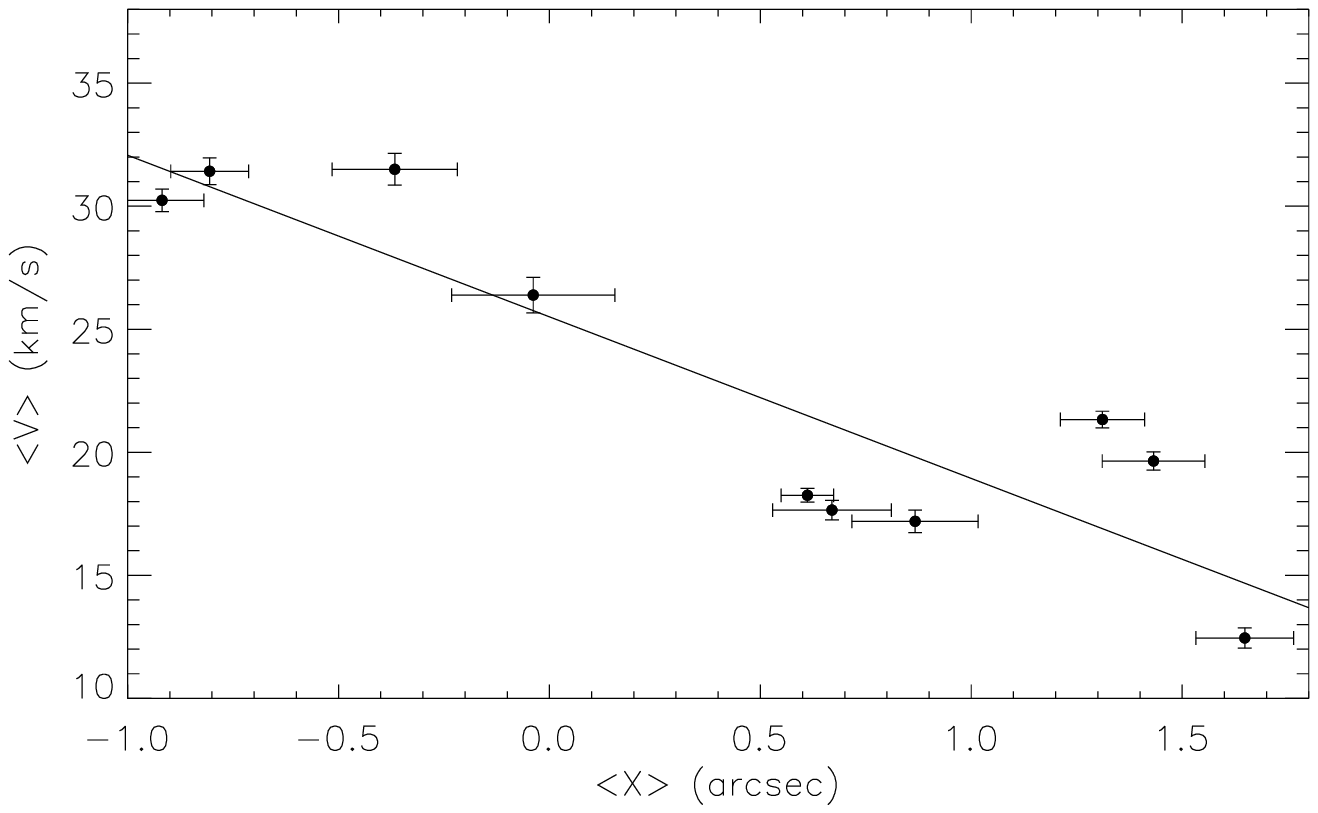}\hspace{0.5cm}
\includegraphics[width=8cm]{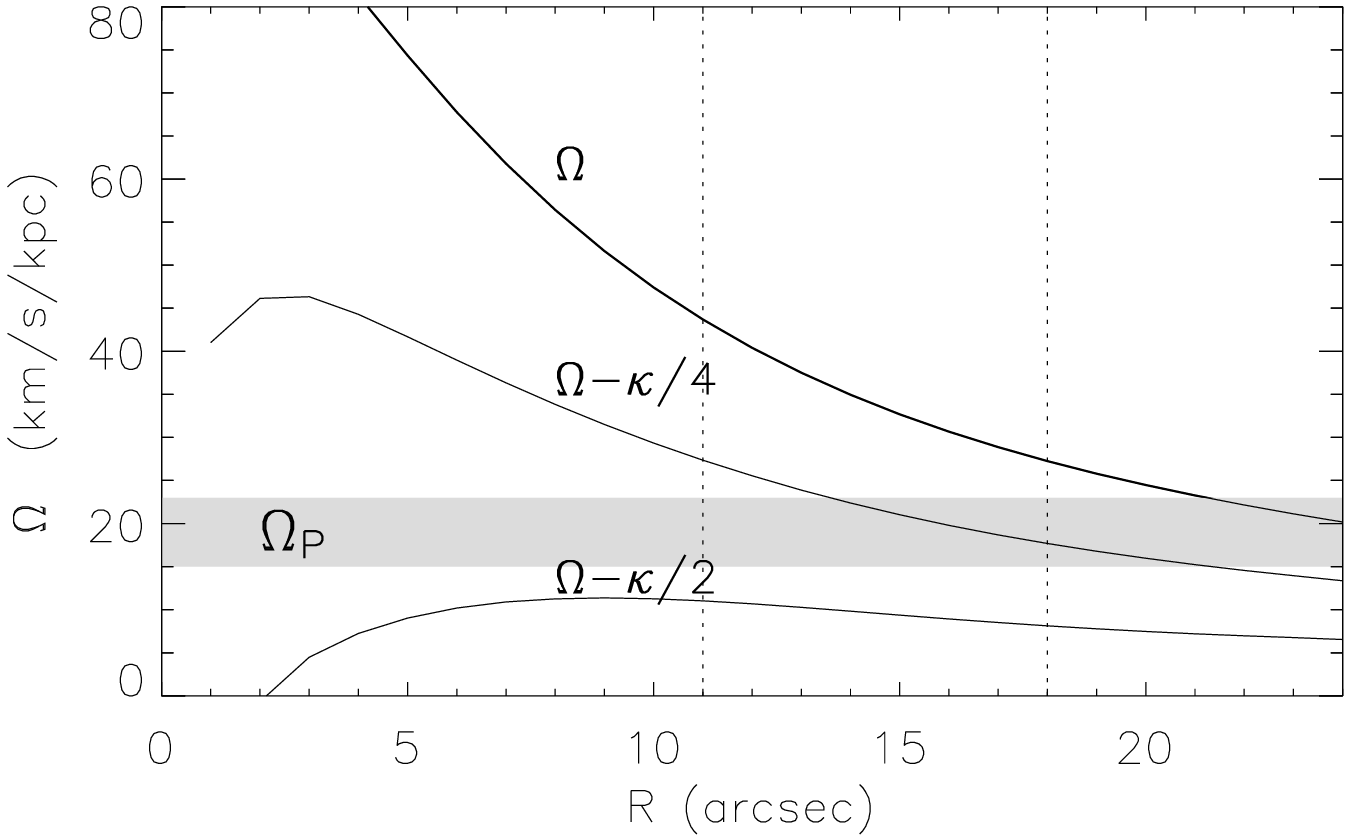}
\caption{Analysis of the inner kinematics of NGC~6104. (a) Radial
variations in the position angle of the kinematic axis. The
dotted line indicates the orientation of the line of nodes and
the solid line indicates the location of the central bar. (b) The
observed rotation curve of the galaxy; the solid line indicates a
polynomial fit. (c) Determination of the bar pattern speed by the
Tremaine-Weinberg method. (d) The diagram of characteristic
frequencies. The gray color marks the range of possible bar
pattern speeds ($\Omega_p$) the dotted lines indicate the maximal
and minimal ring radii.}
\end{figure*}

The TW method was used for measuring $\Omega_p$ in the stellar
components of SB0 galaxies (Aguerri et al. 2003; Debattista 2003)
and in the molecular disks of late-type galaxies (Zimmer et al.
2004; Rand and Wallin 2004). Here, we do not consider the
possibility that the validity conditions of the TW method are
satisfied in  concrete objects as well as the possible error
sources, referring the reader to the above-mentioned papers.

Hernandez et al. (2005) applied the TW method to the FPI ionized
gas velocity field. However, in our point of view, the reasoning
of these authors regarding the applicability of the TW method to
the ionized gas distribution is not convincing, primarily because
there is no simple relation between the gas density and its
surface brightness in \Ha. In the case of NGC~6104, however, it has
already been pointed out that the gas and stellar velocity fields
are similar, the difference lies mainly only in the rotation
amplitude, while the behavior of $PA_{kin}$ along the radius is
the same. Therefore, to a first approximation, the \Ha velocity
field may be assumed to be equivalent to the stellar velocity
field and, if we use the continuum brightness as a weight factor,
then we can try to apply the TW method to our FPI observations. Of
course, such a somewhat liberal treatment of the TW method
requires more careful reasoning, but preliminary results show that
the values of $\Omega_p$ obtained by the method described above
for several galaxies observed with the 6-m telescope are in
satisfactory agreement with other methods for estimating the
positions of resonances (Afanasiev and Poltorak, private
communication). Thorough substantiation and testing the
applicability of the TW method to these data require an individual
study that is outside the frame of this paper.

We took several strips parallel to the line of nodes through the
velocity field and the galaxy continuum image and calculated
$\langle X \rangle$ and $\langle V \rangle$ in each of them. The
strip width and step was 2\farcs1, i.e., 3 pixels. Figure 6c shows
a clear correlation between $\langle X \rangle$ and $\langle V
\rangle$, as follows from Eq. (2). Linear fitting of this
dependence yields the bar pattern speed $\Omega_p =(19 \pm 4)
\km\, \mbox{kpc}^{-1}$ (to within the sign defined by the chosen
direction of the $x$ axis). Here, the uncertainty in choosing the
position angle was assumed to make the main contribution to the
error in the $\Omega_p$ estimation (Debattista 2003). We varied
$PA$ within $PA_0 \pm 5^\circ$, which, in general, slightly
exceeds its formal measurement error, and determined how much
$\Omega_p$ changed in this case.

Figure 6d shows the radial dependence of the angular
velocity $\Omega$ and the $\Omega\pm \kappa/2$ and $\Omega -
\kappa/4$ curves, where $\kappa$ is the epicyclic frequency. In
calculating $\Omega$ and $\kappa$, we used a seventh-degree
polynomial fit to the galactic  rotation curve to avoid any drops
in the derivative $d\Omega/dr$. An examination of the constructed
resonance diagram shows that the adopted bar pattern speed
intersects the $\Omega - \kappa/4$ curve at a distance of $17 \pm
3''$. That is the 1:4 resonance coincides with the ring, within
the measurement error limits, as follows from the theory.
Moreover, at this bar pattern speed, the $\Omega_p = const$ line
does not intersect the $\Omega- \kappa/2$ curve. In other words,
the bar hasn't inner Lindblad resonance (ILR) where the
radial gas motions toward the center cease (Combes and Gerin
1985), so gaseous clouds can reach the AGN, providing fueling for
it.

It should be emphasized that the main conclusion that follows
from Fig. 6d, the absence of ILR, can also be reached without
using the TW method. It will sufficient to assume that the ring is at
the 1:4 resonance and this is the most likely assumption, as we
noted at the beginning of this section. It then follows from the
diagram of characteristic frequencies that the bar pattern speed
is $\Omega_p = 17-27 \km\, \mbox{kpc}^{-1}$, which matches the results of
the TW method, within the error limits.

\section{Examination of the galactic environment.}

Based on images from the 6-m telescope, we have detected
asymmetric filaments of low surface brightness in the outer
regions of NGC~6104 located at apparent galactocentric distances of
25-45 kpc. In contrast to spiral density waves, these structures
are clearly non-stationary and must dissipate in a time of the
order of several periods of disk rotations at a given radius.
Under the assumption of a flat rotation curve, we find the
characteristic lifetime of the filaments is $0.6- 1.0$ Gyr, which
imposes a constraint on the maximum age of the event that
generated them. As we have already noted above, these structures
resemble in shape the result of the disruption of a companion
galaxy by tidal forces. Moreover, they are very similar to the
filaments near the disk galaxy NGC~2782. In the latter object, radio
data on the HI distribution convincingly prove that a merger with
a companion with a mass of $\sim1/4$ of the galactic mass took
place (Jogee et al. 1999). Besides, after the deprojection of the
NGC~6104 image to the face-on position and the subtraction of a
two-dimensional disk model, it became clear that the outer shells,
which are clearly seen in the images in all three color bands, are
a single structure  warps around the disk of NGC~6104, confirming
once again the hypothesis of a disrupted companion. In general,
however, an alternative explanation is also possible -- these
structures are tidal in nature and resulted from a perturbation of
the galactic disk during a close passage of a companion without
its disruption. The companion must then be still close enough. Let
us consider this case in more detail.

NGC~6104 belongs to the cluster Zwicky 1615.8+3505,
also known as N45-389 in the X-ray field. According to Tomita et
al. (1999), this is a dynamically young poor cluster 4 Mpc in
diameter. Although the cluster is fairly poor, the contribution
from the possible interactions of NGC~6104 with its other members
cannot be disregarded.

Four extended objects that can play the role of dwarf companions
are seen in the POSS2 image  in the immediate vicinity of NGC~6104
(within three galactic optical diameters). These are marked by
numbers in Fig. 3a. Since the distances to them are unknown, we
obtained spectra for all four objects. Analysis of the spectra
containing both absorption and emission features shows that these
are relatively distant galaxies (with redshifts of 0.096, 0.220,
0.316, and 0.097, respectively) and their proximity to NGC~6104 is
a projection effect.

Another candidate companion is the elliptical galaxy CGCG 196-022
(CGCG 1615.0+3550), at $4'$ from NGC~6104, and, according to
Tomita et al. (1999), its luminosity is a factor of three lower.
The radial velocity difference between NGC~6104 and CGCG 196-022
is $205 \km$ (from NED). Since the velocity dispersion of the
cluster galaxies is high ($584 \km$ in Tomita et al. 1999), we
estimate a minimal time of possible encounter between these two
galaxies, 0.2 Gyr ago, which is considerably shorter than the
characteristic lifetime of the filaments. Thus, if this galaxy had
a specific trajectory, it could encounter NGC~6104 and produce the
mentioned above tidal structures; but the galaxy CGCG 196-022
itself must also be perturbed appreciably, since the masses of the
two galaxies are similar. However, analysis of our deep images
obtained with the 6-m telescope for CGCG 196-022 has revealed no
perturbations of its isophotes, except an elongated feature on the
periphery of the galaxy southwest of the nucleus, which is also
clearly seen in Fig. 3a. We obtain a spectrum of this feature. In
the range $4020-5600$\AA\, (see Table 2), we detected only one,
but fairly seen intense emission feature at a wavelength of
4362\AA. For moderate redshifts, such situation is possible only
if we observe the [OII]$\lambda3727$ line; thus, the velocity of
this structure is $\sim51200 \km$ and it is a distant background
edge-on galaxy.

 Thus, although the only one candidate galaxy that
could affect appreciably the gravitational potential of the disk
in NGC 6104 is CGCG 196-022, there is no conclusive evidence for
their recent interaction. On the other hand, since CGCG 196-022 is
an elliptical galaxy, the tidal distortions in it may be
considerably smaller than those in NGC~6104, a late-type disk
galaxy. However, the appearance of the external structures more closely corresponds to the
disruption of a dwarf companion (a single strongly warped
structure; see above for the analogy with NGC~2782) than to the
tidal tail from the interaction with a galaxy of comparable mass.
In the latter case, one might expect the presence of a bridge
between the galaxies. This is further confirmation that the outer
filaments in the image of NGC~6104 are the result of the disruption
of a companion rather than the tidal perturbation in the disk.

\section{Conclusions.}

We investigated in detail the active galaxy NGC~6104 and its
environment by means of the integral-field spectroscopy. We
constructed the ionized gas and stellar velocity fields and
studied the contributions from various mechanisms to the gas
ionization in the inner ($r < 5$ kpc) region using diagnostic
diagrams. Based on surface photometry, we refined the
morphological type of the galaxy, Sc. During our work, we were
able to analyze the peculiarities of the structure of NGC~6104 on
various spatial scales.

\begin{enumerate}
\item In the innermost region of the galaxy, we suspected
the existence of a jet from the AGN with a size of no more than
several hundred parsecs. To be more precise, we found effects
that could be explained in terms of the interaction between a
radio jet and the ambient interstellar medium: an excess of
``blue'' velocities in all lines of ionized gas (up to $40 \km$
along the line of sight), a peculiar appearance of the [OIII]
velocity field, and the possible contribution from shocks to the
ionization of the ambient gas that follows from diagnostic
diagrams. To our regret, there are no radio observations
of NGC~6104 with an angular resolution high enough to detect this
radio jet.

\item At larger galactocentric distances (1-5 kpc),
we identified radial motions of ionized gas along the bar toward
the nucleus with velocities of $\sim50 \km$. Diagnostic diagrams
and the observed deceleration of rotation in the [NII] line
relative to \Ha point to an appreciable contribution from the
shock fronts at the bar edges to the ambient gas ionization. We
measured the bar pattern speed by the Tremaine-Weinberg method.
Since the inner Lindblad resonance was shown to be absent in the
bar, the radial gas motions can reach the AGN, fueling its central
engine.

\item We considered the circumnuclear ring and showed that 70\% of
all the current star formation in the galaxy
($SFR\approx7\,M_\odot yr^{-1}$) proceeds in it. The ring is at
ultraharmonic (1:4) resonance of the bar. We noted a distorted
shape of the ring (first of all, a displacement of the center
relative to the nucleus); it could be that this asymmetry was
produced by a recent interaction with a relatively massive
companion.

\item Finally, in the outermost regions of the galaxy at
galactocentric distances of 25-45 kpc, we detected extended
asymmetric filaments of low surface brightness that are clearly
seen in all three bands, BV\Rc. Similar shells, the so-called
ripples, can be formed in galaxies during a close passage of a
companion galaxy or during accretion of dwarf galaxies (Schweizer
and Seitzer 1988; Kormendy 1989). Analyzing the images from the
6-m telescope, we found in the residual images (after the
subtraction of a disk model) that these are not two different
shells, but a single structure that resembles most closely a
companion disrupted by tidal forces.

\end{enumerate}

 So, is a relation between the large-scale outer structures (the
remnants of the merged satellite) and such inner features of the
galaxy as the radial motions toward the center and existence of an
AGN? Here, two alternatives are possible: either the ring and the
bar existed in the galactic disk before the interaction with a
companion or they emerged already during this interaction. In any
case, the interaction leads to a mass redistribution in the
galaxy, primarily to a higher mass concentration in the central
(one kiloparsec size) region (Combes 2001). If the bar and the
ring already existed in the galaxy, then such a redistribution
could increase the rate of gas inflow through the bar into the
AGN. A number of peculiarities in the shape of the ring noted
above (asymmetry, displacement of the center) can then be
explained by distortions of the gravitational potential via the
interaction. In this case, the merger with a companion must have
taken place much later than $\sim1.0$ Gyr determined in the
previous section, since the dynamical times at $r = 5$ kpc are an
order of magnitude shorter than those at $r = 50$ kpc and the ring
had not yet been perturbed and relaxed.

The more courageous assumption that both the bar and the ring are
the result of interaction with a companion is supported by the
following circumstantial evidence. We see from Fig. 3c that the
concentration of dust lanes in the HST image is highest in the
northwestern half of the galaxy. If the dust lies in the disk
plane, then this half must be closer to the observer. Since it
follows from the velocity field that the northeastern edge of the
galaxy recedes from us, then we find that the galactic disk
rotates counterclockwise. However, the spirals in the disk are
then the leading ones; this is a very rare situation that is
possible only in interacting systems in which the spiral density
wave is generated by a companion (Pasha 1985). We then find that
the entire spirals-ring-bar structure resulted from the
gravitational tidal effect of a companion (the situation in which
the bar and the spirals rotate in opposite directions seems a
short lived one and breaks up after several bar rotations). This
scheme is very attractive, but note that the dust distribution in
the circumnuclear region is still very complex and further
arguments for such a spatial orientation of the galactic disk are
required.

Note once again a number of analogies between NGC~6104 and
NGC~2782, which was studied in detail by Jogee et al. (1999). Both
objects are disk spiral barred galaxies with their outer isophotes
perturbed by a recent interaction that ended with the merging with
a less massive companion. In both cases, outer filaments with a
similar structure (the remnants of the disrupted companion) are
observed. In NGC~2782 this interaction led to the concentration of
gas in the inner region, the formation of a bar from molecular
gas, and a powerful starburst. In NGC~6104, we also see gas flows
toward the center where the AGN is located. Although it is not yet
clear why similar causes led to the formation of different types
of nuclear activity (a Seyfert nucleus and a circumnuclear
starburst), a relationship with a recent merging event is
traceable in both objects.

According to the studies listed in the Introduction, which are
based on samples of both Seyfert and normal galaxies, no
statistically significant difference has been found between these
two classes in terms of the presence of companions or traces of
interaction. In most studies, the authors restricted themselves to
searching for traces of large-scale interactions with companions
of comparable mass (major-merging). However, a more detailed study
of the surroundings of galaxies can revise these conclusions; the
apparently single (at first glance) galaxy can be in the process
of merging with a low-mass companion, as we found in our case. A
similar situation is also observed, for example, in the active
galaxy Mrk 315. Previously, it was considered in all papers as a
single galaxy, but deep images from the 6-m telescope (Ciroi et
al. 2005) show that this galaxy simultaneously interacts with two
dwarf companions, one of which is in the process of tidal
disruption. Finally, we must emphasize that the AGN activity  in any galaxy is
not necessarily caused by the presence of a bar or traces of a
recent interaction. We  try to draw attention to the fact that
only a careful analysis of each object using currently available
observational facilities allows the factors that caused its
current activity to be established.

\begin{acknowledgements}
This work is based on observational data obtained from the 6-m SAO
telescope financed by the Ministry of Science of Russia
(registration no. 01-43) and NASA/ESA Hubble Space Telescope data
retrieved from the Space Telescope Science Institute archive
operated by the Association of Universities for Research in
Astronomy Inc. under contract with NASA (NAS 5-26555). We used the
NASA/IPAC Extragalactic Database (NED) operated by the Jet
Propulsion Laboratory of the California Institute of Technology
under contract with the National Aeronautics and Space
Administration (USA). We specially thank T.A. Fatkhullin, who
performed calibration observations at the Zeiss-1000 telescope
(SAO). The work was supported in part by the ``Nonstationary
Objects in the Universe'' Program of the Ministry of Industry and
Science. A.V.Moiseev is also grateful to the Russian Sciences Support Foundation
 and the Russian Foundation for Basic
Research (project no. 05-02-16454) for financial support of our
study.
\end{acknowledgements}

\textit{Translated by V.~Astakhov}

\end{document}